\documentclass[preprint,3p]{elsarticle}

\usepackage[utf8]{inputenc}
\usepackage[english]{babel}

\usepackage{lineno,hyperref}
\modulolinenumbers[5]

\usepackage[capitalize]{cleveref}

\usepackage{soulutf8}

\soulregister\cite7
\soulregister\ref7
\soulregister\cref7
\soulregister\pageref7

\crefname{algorithm}{Algorithm}{Algorithms}
\crefname{example}{Example}{Examples}
\crefname{part}{Part}{Parts}
\crefname{table}{Table}{Tables}
\crefname{figure}{Figure}{Figures}
\crefname{chapter}{Chapter}{Chapters}
\crefname{section}{Section}{Sections}
\crefname{appendix}{Appendix}{Appendices}
\crefname{lstlisting}{Listing}{Listings}

\usepackage{paralist}

\usepackage[dvipsnames]{xcolor}

\newcommand\addedtextRR[1]{\textcolor{Green}{#1}}

\usepackage{listings}
\crefname{lstlisting}{Listing}{Listings}

\usepackage{soulutf8}
\usepackage{xcolor}

\usepackage{subcaption}

\usepackage{listings}

\definecolor{codegreen}{rgb}{0,0.6,0}
\definecolor{codegray}{rgb}{0.5,0.5,0.5}
\definecolor{codepurple}{rgb}{0.58,0,0.82}
\definecolor{backcolour}{rgb}{0.95,0.95,0.92}

\lstdefinestyle{mystyle}{
    backgroundcolor=\color{backcolour},
    commentstyle=\color{codegreen},
    keywordstyle=\color{magenta},
    numberstyle=\tiny\color{codegray},
    stringstyle=\color{codepurple},
    basicstyle=\fontsize{7}{9}\ttfamily,
    breakatwhitespace=true,
    breaklines=true,
    captionpos=b,
    keepspaces=true,
    numbersep=5pt,
    showspaces=false,
    showstringspaces=false,
    showtabs=false,
    tabsize=4,
    frame=tblr,
    framerule=0pt,
    columns=flexible
}
\usepackage{soul}

\title{Local Features: Enhancing Variability Modeling in Software Product Lines}

\begin{document}

\author[lbd]{David de Castro}
\ead{david.decastro@udc.es}
\author[lbd]{Alejandro Corti\~nas}
\ead{alejandro.cortinas@udc.es}
\author[lbd]{Miguel R. Luaces}
\ead{luaces@udc.es}
\author[lbd]{Oscar Pedreira\corref{mycorrespondingauthor}}
\ead{oscar.pedreira@udc.es}
\author[lbd]{Ángeles Saavedra Places}
\ead{asplaces@udc.es}

\address[lbd]{Universidade da Coru\~na, Centro de Investigación CITIC, Database Lab. \\ Elviña, 15071, A Coru\~na, Spain}
\cortext[mycorrespondingauthor]{Corresponding author}


\begin{abstract}
\noindent {\em Context and motivation:} Software Product Lines (SPL) enable the creation of software product families with shared core components using feature models to model variability. Choosing features from a feature model to generate a product may not be sufficient in certain situations because the application engineer may need to be able to decide on configuration time the system's elements to which a certain feature will be applied. Therefore, there is a need to select which features have to be included in the product but also to which of its elements they have to be applied.

\noindent {\em Objective:} We introduce \emph{local features} that are selectively applied to specific parts of the system during product configuration.

\noindent {\em Results:} We formalize local features using multimodels to establish relationships between local features and other elements of the system models. The paper includes examples illustrating the motivation for local features, a formal definition, and a domain-specific language for specification and implementation. Finally, we present a case study in a real scenario that shows how the concept of local features allowed us to define the variability of a complex system. The examples and the application case show that the proposal achieves higher customization levels at the application engineering phase.
\end{abstract}
  

\maketitle

\section{Introduction}
\label{sec:intro}

Software Product Lines (SPL) support the semi-automatic development of families of software products. The products in the family are built from a set of common {\em core assets} and share many {\em features}, although they differ in others. The SPL development paradigm structures the development process in two phases: the {\em domain engineering} phase and the {\em application engineering} phase. 
The goal of the {\em domain engineering} phase is to analyze, design, and build the components of the product family and the SPL platform. A key activity of this phase is the analysis and modeling of variability, that is, the identification of the features that can be present in the products of the family and the relationships between them. The goal of the {\em application engineering} phase is the configuration and generation of a specific product. The application engineers select which features must be present in that product and the SPL platform generates it by adapting and combining the core assets according to the selected features \cite{weiss99,Pohl2005}. SPLs mean a great advance in the development of software product families since the automatic generation of these products led to a considerable reduction of development costs and an increase in product quality. This is the reason why there are more and more SPLs orientated to very different application domains~\cite{Weiss2006}, from the generation of software for airplanes~\cite{boeing-hall-of-fame}, software for IoT devices~\cite{Iglesias201964}, web portals~\cite{Trujillo2007a}, or virtual stores~\cite{Rincon201571}, just to name a few examples.

A {\em feature} represents a characteristic of a system and can be related to a functionality, a design decision, or other elements of the system \cite{Pohl2005,apel2009overview,apel2016feature,Benavides2010}. The variability of a SPL is usually described in a feature model, a hierarchical tree that represents the features that may or may not be present in any of the products of the family, and the relationships between them \cite{Pohl2005,apel2016feature}. In the domain engineering phase, feature models allow domain engineers to define the variability of the SPL, which is essential in the design of the different components. In the application engineering phase, the SPL platform allows the application engineer to generate a product of the family by selecting the features that must be present in that product. In this way, it is possible to generate software that contains only those functionalities that are necessary. For example, although all e-commerce stores are very similar and share many features, there are some functionalities that may differ, such as the payment methods: some allow payment by credit card, some with PayPal, and some with both options. These payment methods are features of the product family and can be present in some products but not in others.

Managing variability with feature models is insufficient in many scenarios because the level of customization they allow can be limited for complex systems. This is why extensions to the feature model have been developed to improve variability modeling. 
Among these extensions are feature models with attributes~\cite{Czarnecki2002,Czarnecki2005b,Benavides2005,Batory2006,6030048}, which allow adding additional information to a feature; or cardinality-based feature models~\cite{Riebisch2002,Czarnecki2002,Czarnecki2004,Czarnecki2005}, which allow defining cardinalities on subtrees of the feature model, that can then be replicated according to those cardinalities, therefore creating several instances of the same feature. Both extensions considerably improve the level of expressiveness of the feature models.

\vspace*{4pt}
\noindent {\em Problem Statement}

In previous real projects where we applied the SPL paradigm, we encountered certain scenarios that highlighted the necessity for a more detailed level of product specification and customization beyond what traditional feature models could offer. 
The problem we have identified is that in certain situations, it is desirable to be able to express that  some functionality of a product is applied only to specific parts of it, and these {\em parts} cannot be identified during the domain engineering phase but must be identified during the application engineering phase. Hence, we found that it is not sufficient to decide whether a feature is included or not in the product; we also need to specify to which elements of the system the feature will be applied. In other words, the features selected in the product configuration cease to be global (for the entire product) and become local (for specific parts of the product). Next, we present some examples to illustrate the motivation for this research problem.

\begin{figure}
    \centering
    \includegraphics[width=0.6\textwidth]{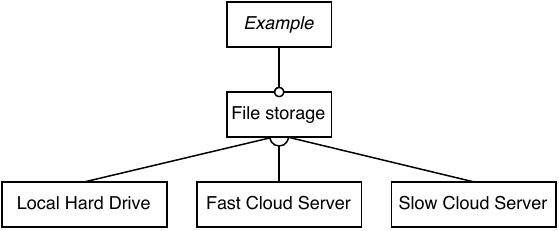}
    \caption{Excerpt of the example feature model of a file storage system}
    \label{fig:example-feature-model-1}
\end{figure}

\vspace*{4pt}
\noindent {\em Motivating examples}

{\em File storage.} 
Many software systems must manage a large number of files. 
Consider, for example, a SPL for managing a family of document management systems for administrative processes in the public administration (with functionalities such as directory management, document storage, document review and approval workflows, etc.)
The data model of this software product line would have a large number of classes in which one of their properties would be a file (e.g., the employee record would have a profile photography, the regulation class would have the official gazette file, the citizen complaint class would have a collection of supporting evidences, etc.). 
This SPL could also have three alternatives for storing the files: storing them on a local hard drive, storing them on a cloud server with fast access, and storing them on a cloud server with cheaper but slower access time. 
Each of these alternatives would be represented as a feature in the feature model of the SPL (e.g., \texttt{Local Hard Drive}, \texttt{Fast Cloud Server}, and \texttt{Slow Cloud Server}, in the feature model excerpt shown in \cref{fig:example-feature-model-1}). 
Let us consider an additional requirement: since there is a substantial number of classes containing file attributes, potentially reaching the order of hundreds, which is justifiable in the context of an extensive document management software product family, we aim to have the capability to individually select the storage type for each file, providing granular control over the storage mechanism. 
The application engineer may need to select different storage options for different classes of files attending to the product's requirements and to optimize the access-time/cost trade-off. For example, in the case of a profile photograph, it is essential to serve it from a fast cloud server due to the significance of speed for optimal user experience. On the other hand, a gazette file, which may not require rapid access, can be served from the local hard drive. Lastly, citizen complaint evidences, being infrequently accessed, can be efficiently served from a slow cloud server.

\begin{figure}
    \centering
    \includegraphics[width=0.9\textwidth]{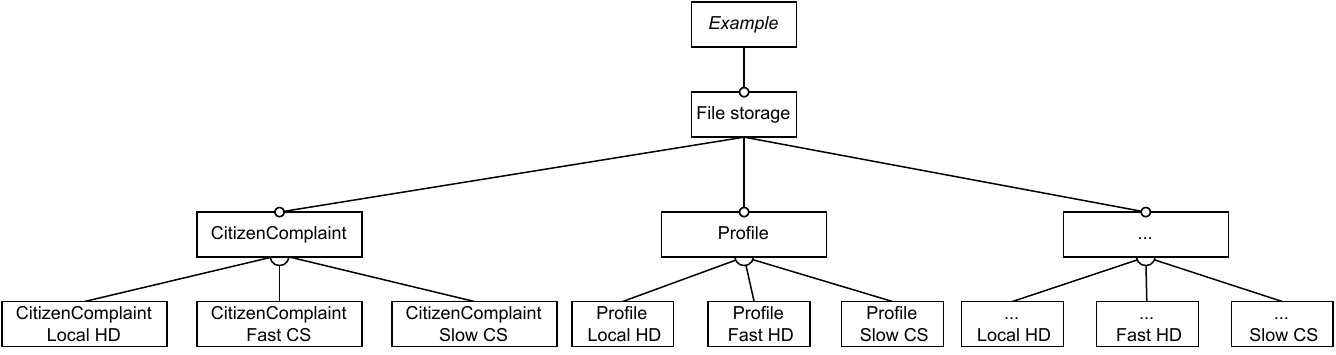}
    \caption{Excerpt of the example feature model of a file storage system SPL supporting granular control}
    \label{fig:example-feature-model-ii}
\end{figure}

\begin{figure}
    \centering
    \includegraphics[width=0.5\textwidth]{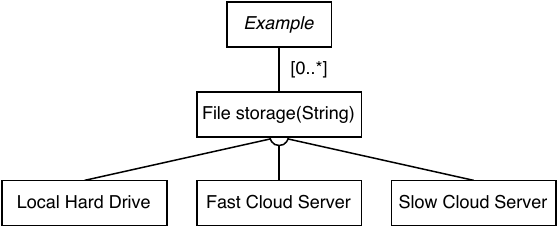}
    \caption{Excerpt of the example cardinality-based feature model of a file storage system SPL}
    \label{fig:example-feature-model-iii}
\end{figure}

\begin{figure}
    \centering
    \includegraphics[width=0.9\textwidth]{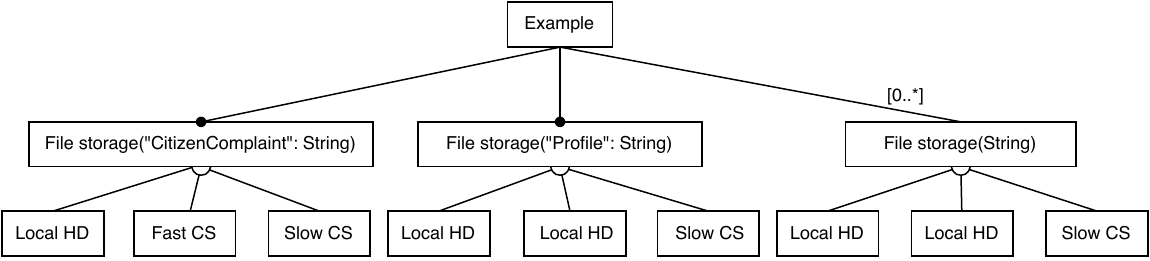}
    \caption{Specialization of the example in \cref{fig:example-feature-model-iii}}
    \label{fig:example-feature-model-instance-iii}
\end{figure}

\begin{figure}
    \centering
    \includegraphics[width=0.5\textwidth]{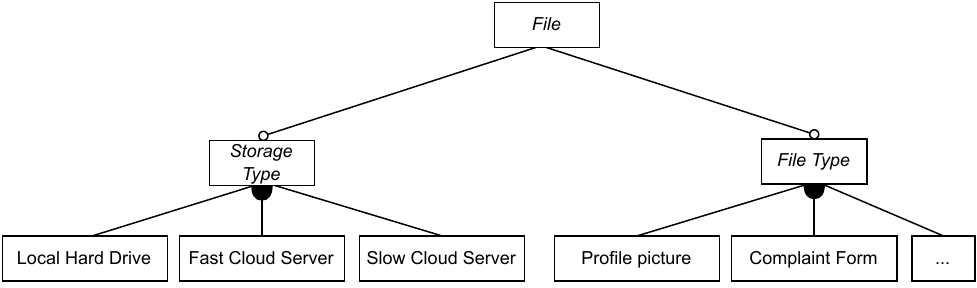}
    \caption{Excerpt of the example solution with features binding for a file storage system SPL.}
    \label{fig:example-feature-model-iv}
\end{figure}

\begin{figure}
    \centering
    \includegraphics[width=0.3\textwidth]{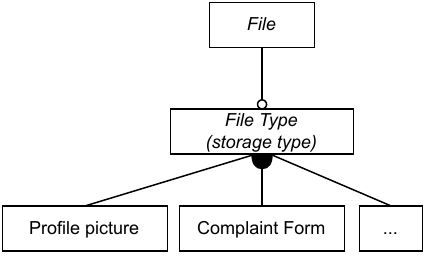}
    \caption{Excerpt of the example solution with features representing file types with attributes for defining the storage alternative.}
    \label{fig:example-feature-model-v}
\end{figure}

There are already ways of addressing this scenario using the current state of the art:
\begin{enumerate}
    \item[(i)] An approach would be to use a feature model similar to the one shown in \cref{fig:example-feature-model-1}. The application engineer would only be able to select one of the three features for a specific product, and it would be applied uniformly to all files across all classes. As a result, the SPL would not be able to meet the requirement that each file can utilize a different storage system.
    
    \item[(ii)] The second solution  would involve replicating the subtree shown in \cref{fig:example-feature-model-1} as many times as the number of classes for which the \texttt{File Storage} feature can be applied. This would result in a feature diagram with dozens of replicated subtrees (see \cref{fig:example-feature-model-ii}). During the application engineering phase, the engineer would have to indicate for each class which of the three storage alternatives would be applied. We believe that this solution leads to an unnecessarily complex feature diagram, which also necessitates a complex product configuration during the application engineering phase.

    \item[(iii)] The third solution would involve applying cardinality-based feature models. The \texttt{File Storage} feature would be cloned as many times as there are classes with files in the data model (see \cref{fig:example-feature-model-iii}, with the notation of \cite{Czarnecki2005} for the cardinality-based feature model). Each clone of the subtree would be associated with a class of the model using an attribute in the \texttt{File Storage} feature (see \cref{fig:example-feature-model-instance-iii}). While managing the tree is simpler than in the previous case, during the application engineering phase engineers would still have to indicate the choice again for each class in the data model. Furthermore, the association between the features and the classes to which they will be applied is established as an attribute of the feature, which weakens such an association. 

    \item[(iv)] Another alternative is shown in \cref{fig:example-feature-model-iv}. Our feature model could have a root feature \texttt{File} with two mandatory children, \texttt{Storage Type} and \texttt{File Type}. The feature \texttt{Storage Type} feature would have three children, each representing a file storage alternative (local HD, fast cloud server, and slow cloud server). The feature \texttt{File Type} would have as many children features as file types we would need to store in the system (such as profile picture, for example). The decision of which storage option would be applied to each file type would be decided at the product's configuration time by binding child features of \texttt{File Type} with children features of \texttt{Storage Type}. Another possibility would be establishing the relationships between the file types and the applied storage types through cross-tree constraints in the feature model. The main drawback of this solution is that it would imply incorporating all the file types of a particular product into the feature model as features that may not be relevant to other products. This would generate an artificially complex feature model since it would have to incorporate elements that are not necessarily relevant features. In this sense, the feature model would become something like a supermodel that incorporates elements already defined in other system models, which we believe would be a bad design approach violating the principle of separation of concerns. In addition, we would still need to formalize how the binding between the two groups of features would be modelled or defined.

    \item[(v)] Finally, another possible alternative would be having in our feature model an abstract feature \texttt{File Type} with as many children as types of files we would need to store in the system (again, examples include profile picture, gazette, etc.) These features could have an attribute \texttt{storageType} that would specify if that file is stored in a local hard drive, fast cloud server or slow cloud server (see \cref{fig:example-feature-model-v}). This alternative shares one significant drawback with alternative (iv), that is, we would be incorporating into the feature model features that represent the types of files of a particular product that may not be relevant for other products. In addition, the specification of the storage type of the files would be expressed in a weak way, since these storage alternatives would no longer appear as features in the feature model (and we consider they are features that should appear in the feature model).
\end{enumerate}

We believe that these five solutions can be improved. In the case of option {\em (i)}, the solution falls short of meeting the requirements during the application generation phase, as it does not allow the engineer to configure which alternative would be applied to each class. Regarding solutions {\em (ii)} and {\em (iii)}, we find them to be complex, as the feature model repetitively includes the same subtree numerous times, which could be avoided. Additionally, the engineer would need to indicate the chosen storage alternative for each class, even if all classes use the same alternative and only one of them uses a different one. In the case of solution {\em (iv)}, the feature model would have to incorporate elements already defined in the data model as features, resulting in an artificially complex feature model. Finally, in the case of solution {\em (v)}, the features representing the variability regarding storage options would disappear from the feature model (something we consider a bad design), and the relationship between the file types and the applied storage option would be defined in a weak way.

Scenarios similar to the one presented in the previous example can emerge when features represent transversal elements of the software, as exemplified in the following two examples.

{\em Access Logging.} A common requirement for many software systems is to maintain a log of who (and when) performed modifications on the data (in some cases, it is also necessary to keep the original and modified data). This requirement can be included as a functionality in the feature model. This way, the application engineer would have the possibility to specify for each generated product whether the functionality is included or not. However, in most cases, the application engineer would not want this functionality to apply to all data in the system but only to those for which user access auditing may be required. If the application engineer were to apply the functionality to all data, it would entail storing and managing logs that the application engineer does not actually need, which we consider a poor design decision. Therefore, in this example, the application engineer would desire the ability to include the functionality in the product but also to specify which data in the system it will be applied to. The situation is similar to the previous case. The application engineer wishes to select a functionality but does not want it to be applied to all elements in the data model; instead, they want to choose specific ones.

{\em Data Export.} Many enterprise applications allow users to export data in formats like CSV for certain classes in the data model. Data export in CSV can be modeled as a feature in a feature model. However, similar to the previous examples, we may not want this functionality to be available for all classes in the system's data model, but only for those that are required based on the specific software product's requirements.

\vspace*{4pt}
\noindent {\em Definition: Global and Local Features}

We faced scenarios similar to these ones in real projects, which led us to propose the concepts of {\em global features} and {\em local features}. Our proposal includes defining software variability at two levels: {\em global}, which comprises features that apply to variation points defined in the domain engineering phase and {\em local}, which comprises features that apply to variation points defined in the application engineering phase. A global feature retains the semantics found in current feature models; it represents features that can be present in any of the products within the product family. On the other hand, a local feature in a software product line is a functionality that, when selected for inclusion in a product, will be applied solely to specific elements defined in other system models. That is, during the domain engineering phase, the engineer determines that some features may be applied to certain elements of the system models. During the application engineering phase, the engineer must associate these features with the specific elements of the system (specified in other system models) to which they will be applied. Our proposal includes using multimodels to specify the binding ~\cite{siegmund2020configuration} between local features and the elements specified in other system's models. Thus, the formality of the proposal is improved with respect to other alternatives such as using a configuration file at the application engineering phase.

\begin{figure}
    \centering
    \includegraphics[width=0.7\textwidth]{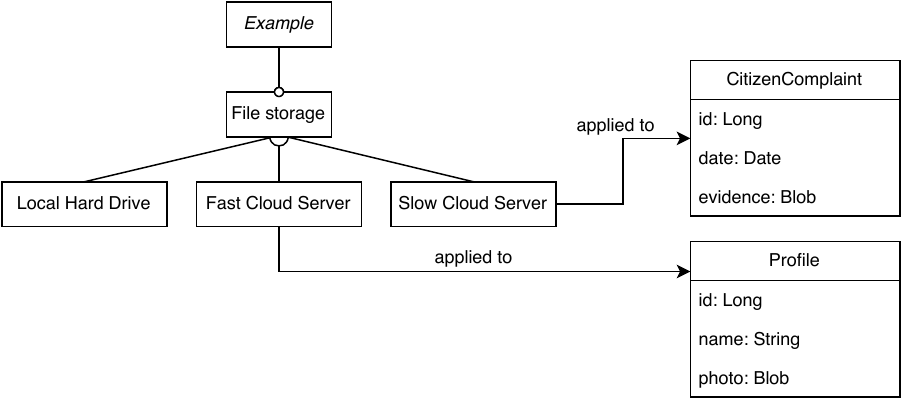}
    \caption{Example of application of local features to a public administration system}
    \label{fig:example-feature-model-2}
\end{figure}

\cref{fig:example-feature-model-2} shows how the previous example would be solved with the concept of local features. In the domain engineering phase, the feature model would specify that there are three alternatives to store files. However, these features are not associated with predefined variation points of the SPL, since, in the excerpt shown in \cref{fig:example-feature-model-2}, the three features \texttt{Local Hard Drive}, \texttt{Fast Cloud Server} and \texttt{Slow Cloud Server} are all local features. In the application engineering phase, the engineer can select a global feature to be applied to all files in all classes by default if no local features are specified. Furthermore, the engineer may select local features to be applied to specific elements of the data model. In the example shown in \cref{fig:example-feature-model-2}, an excerpt from the data model has two classes, \texttt{Profile} and \texttt{CitizenComplaint}. The class \texttt{Profile} is associated with the feature \texttt{Fast Cloud Server}, as it may be necessary to retrieve these files quickly. Similarly, the class \texttt{CitizenComplaint} is associated with the feature \texttt{Slow Cloud Server}, since evidence files require a lot of storage space and are not often accessed. The files in other classes (e.g., the regulation class) are stored as local files because \texttt{Local Hard Drive} was selected as the global feature. In this way, we could select a predefined storage alternative for all classes, but we would be able to customize the storage alternative for specific classes by associating them with local features. This approach leads to simpler feature models and simpler product configurations in the application engineering phase.

In the rest of the article, we develop the concepts of global and local features and how they can be implemented in practice. From the point of view of modeling and specification, we decided to implement global and local features through the concept of multimodels, which has already been used in some previous works as a way to specify the relationships between feature models and other system's models \cite{multimodels-quality}.  From the point of view of the practical specification of global and local features, we propose using a Domain-Specific Language (DSL) that enables the application engineer to define a product specification including the aforementioned relationships between local features and other system elements. We introduced the concept of local features informally in a previous conference paper \cite{decastro22}, in which we showed a specific industrial experience that motivated the need for local features in some SPLs.

Finally, the article presents an application case in a real scenario in which we applied the concept of local features to a SPL for the generation of geographic information systems (GIS), that is, systems in which many of the entities have a geo-spatial component as part of their attributes. This application case shows how the concept of local features allowed us to specify and configure how the features related to data visualization can be applied and adapted to the user visualization requirements in a real system.

The rest of the paper is organized as follows. \cref{sec:background} provides an overview of previous work. \cref{sec:proposal} explains our proposal, conceptually, with a brief example, and shows the changes that had to be made to our SPL's variability models to support it. \cref{sec:case-study} presents a case study to illustrate our proposal with an existing GIS web application, showing an example step by step. Finally, \cref{sec:conclusions} concludes the paper and comments on some ideas for future work.
\section{Background and Related Work} 
\label{sec:background}

\subsection{Software Product Lines and Feature Modeling}

Software Product Lines (SPL) are one of the solutions adopted by the software industry to develop quality software in reduced time. An SPL is a platform that supports the development of a family of software products that share a set of common {\em features} but that can vary in others. To semi-automate the development of one of the software products of the family, an SPL allows the engineer to select the features that should be present in that product. Then, the SPL assembles and adapts a set of {\em core components} to generate that product based on the selection of features. The {\em variability} in the features of the product family is represented in a {\em feature model}, which represents the features of the product family and the relationships between them. This approach tries to bring to software development the production schemes that have been applied in the last century in other industries, such as the production of automobiles, for example. There are many examples of applications of SPL in real scenarios. For example, companies in the aviation sector have SPL oriented towards generating software for their aircrafts~\cite{boeing-hall-of-fame}, and there are SPLs oriented towards generating e-commerce web applications \cite{Rincon201571}.

Feature modeling~\cite{Kang1990} is the \textit{de facto} variability representation for SPLs~\cite{Sousa2016,Benavides2010,Galindo2020,Apel2009a}. 
A feature model is a tree where the features (\textit{end user-visible characteristics of a software system}~\cite{Kang1990}) of a product line are hierarchically structured. Each feature can be decomposed into several sub-features, and they can be mandatory, optional, or alternative features~\cite{Pohl2005}. Every product in the product line is specified by the set of features included in it. 
Besides the relationship between a feature and its sub-features, we can define cross-tree constraints between the features, for example, \texttt{including feature A implies that feature B is also included}. These are the components of what can be considered basic feature models. However, previous works have proposed several extensions to these basic feature models~\cite{Benavides2010,Alferez2019}.

Cardinality-based feature models~\cite{Riebisch2002,Czarnecki2002,Czarnecki2004,Czarnecki2005} allow defining UML-like multiplicities or cardinalities for the features. These cardinalities determine the number of instances of a feature that can be included in a product, and each of these features can include a specific set of sub-features. Extended feature models~\cite{Czarnecki2002,Czarnecki2005b,Benavides2005,Batory2006,6030048} allow to link features with \textit{attributes}; this is, each product can include extra information for a selected feature beyond its own selection. These attributes can be, for example, a number within a specific range, or a string literal, and can be used within the constraints between the features. Both of the approaches have been used together~\cite{Czarnecki2002,Czarnecki2005b,karatas2013} to obtain a flexible and complete variability model. 

However, these extensions to basic feature models only allow for expressing certain variability elements in an SPL. Specific domains require additional models and relationships between them to express additional variability.

\subsection{Multimodel Description of Software Product Lines}\label{subsect:multimodel}

\begin{figure*}
  \centering
  \includegraphics[width=0.8\linewidth]{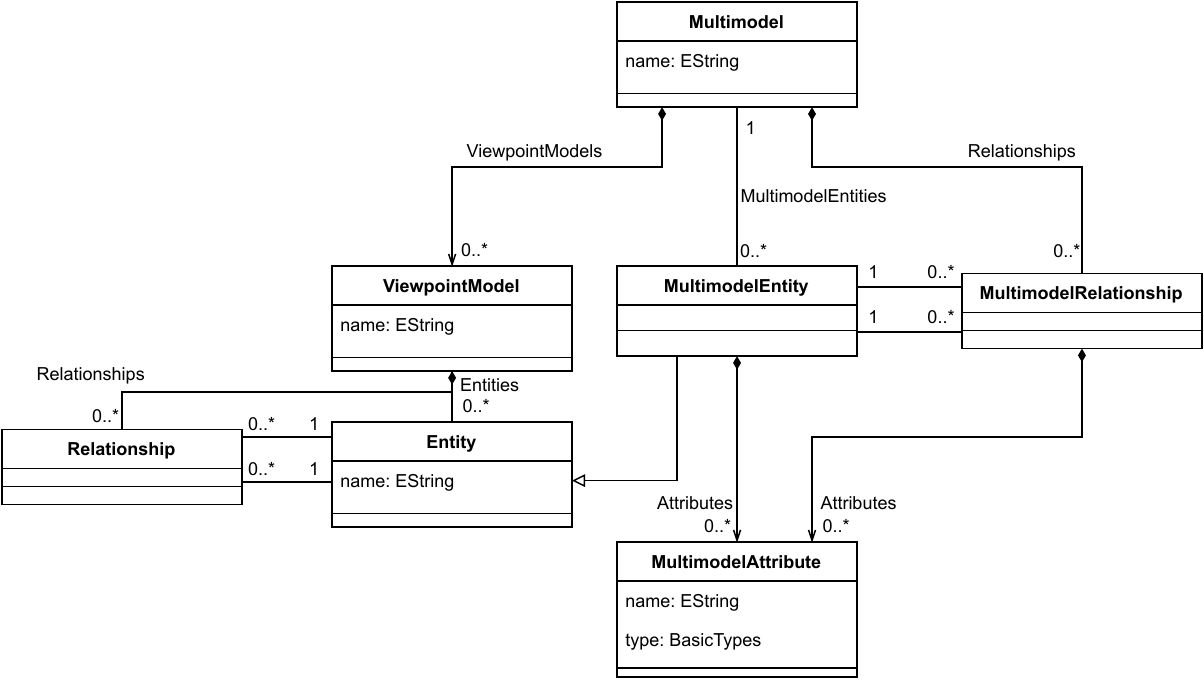}
  \caption{A metamodel of multimodels~\cite{multimodel-metamodel}.}
  \label{fig:multimodel-metamodel}
\end{figure*}

A {\em multimodel} is a set of interrelated models, each of them representing a different viewpoint of the system.  Each {\em viewpoint} is an abstraction of a system part with a specific purpose and is represented through a specific model~\cite{multimodels-def}. In a multimodel description, we can establish relationships between these different viewpoints (models) so that one element of one viewpoint can be related to an element of another. The relationships between elements of different models are defined outside those models, in a specific model of relationships.

\cref{fig:multimodel-metamodel} shows a metamodel of multimodels \cite{multimodel-metamodel}. The left side of the figure contains the classes representing a model of the system. As we can see in the figure, the class \texttt{ViewpointModel} represents a particular model, which is composed of \texttt{Entities} and \texttt{Relationships} between those entities. There will be as many viewpoint models as there are of interest in the domain, and each viewpoint model will, in turn, have as many entities and relationships as necessary to represent the concepts and relationships in the corresponding view model. On the right side of the figure, the rest of the classes represent the concept of multimodel and its components. A \texttt{Multimodel} is composed of \texttt{ViewpointModels} and \texttt{MultimodelRelationships}. A \texttt{MultimodelRelationship} defines the relationship between \texttt{MultimodelEntities}, each of them extending an \texttt{Entity} of a specific \texttt{ViewpointModel}. In addition, the multimodel can also contain \texttt{MultimodelAttributes} associated to either a  \texttt{MultimodelEntity} or a \texttt{MultimodelRelationship}.

Previous works have already used the concept of multimodel to represent information of a software product line. González-Huerta, Insfram, and Abrahão \cite{multimodels-quality} used multimodels to model the relationships between features, elements of the software architecture, and quality attributes of the software product line. This multimodel allows the platform to assess the fulfillment of non-functional quality attributes based on the selection of features and the presence of specific software architecture components in the definition of the product to be generated. The information contained in the multimodel even allows to make recommendations on the features that should be selected to meet those non-functional requirements. 

We have decided to use the multimodel approach instead of CVL or OVM \cite{Pohl2005,metzger2007disambiguating} because we found it to be a straightforward approach to establish relationships between features and other elements. Additionally, it has been used in previous works to establish similar associations, and in terms of practical implementation, it seemed like an easy and manageable option. Moreover, both OVM and CVL can be defined using multimodels, so we are using a more general approach. On the other hand, CVL was dropped due to legal, patent-related issues~\cite{berger2019usage}. Similarly, OVM aims to separate the description of feature variability from the rest of the software artifacts providing a powerful way to handle feature variability and configuration, but it is not specifically designed to represent relationships between models in a direct or formalized manner. Also, in OVM the association between the features and the variation points is defined in the domain engineering phase, and it is not expected to be modified in the application engineering phase.

\section{Local Features in Software Product Lines}
\label{sec:proposal}

In this section, we introduce and formalize the concept of global and local features. We first present the formalization of global and local features using multimodels, and then we provide an example of its application.

\subsection{Definition of Global and Local Features}

\begin{figure*}[htbp]
  \centering
  \includegraphics[width=0.9\linewidth]{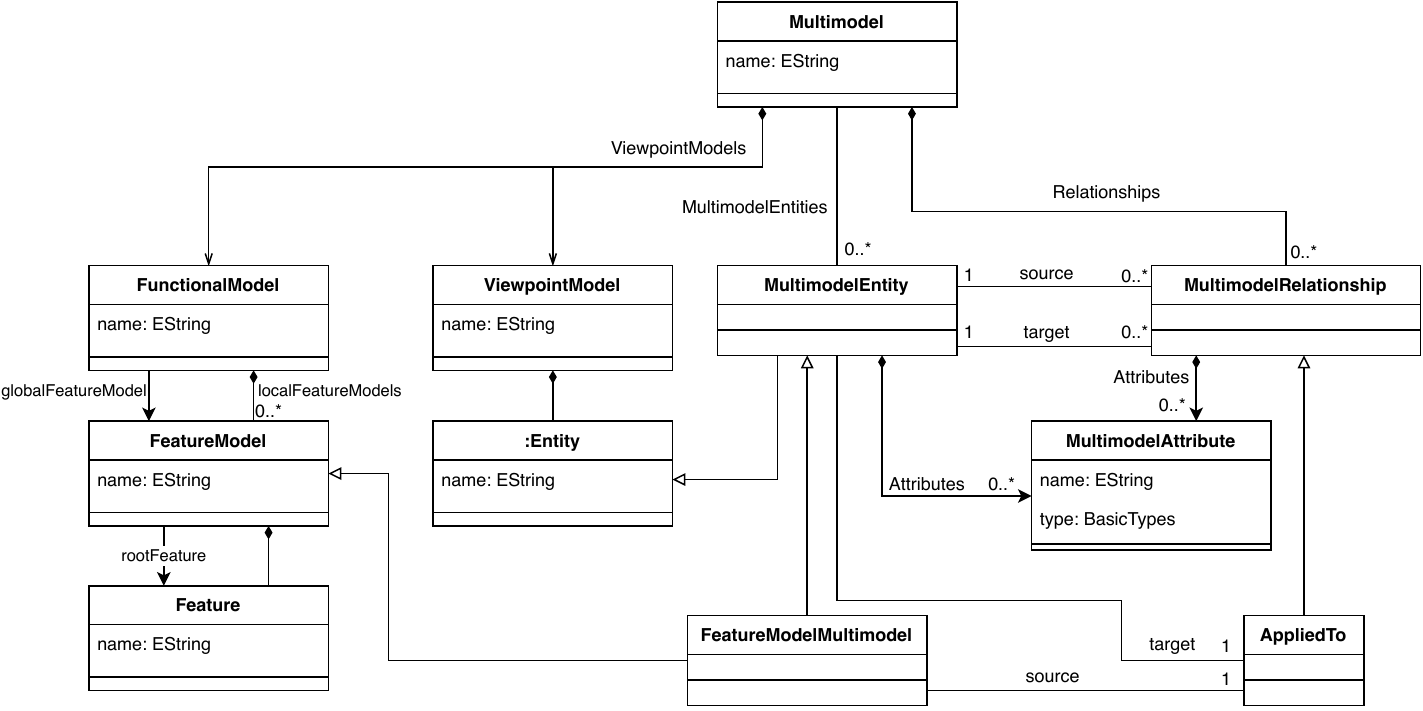}
  \caption{A metamodel of multimodels with the inclusion of local features.}
  \label{fig:multimodel-metamodel-local-features}
\end{figure*}

The definition of global and local features assumes that the products of the family supported by the SPL are defined by multiple models. One of these models must necessarily be a feature model. A {\em local feature} is a feature in a software product line that, if selected to be included in a product, it will be applied only to specific elements defined in different system models, as defined  by the engineer. 
A {\em global feature} has the semantics they have in current feature models. It represents a feature that can be present in any of the family's products. As we have already mentioned, local features will be applied only to specific elements of the system defined in other system's models. The other system's models can represent different viewpoints of the system, such as its architecture, data model, or visualization configuration. 
The relationship between a local feature and the elements in other models to which it will be applied is modeled using multimodels. As we can see in \cref{fig:multimodel-metamodel-local-features}, our multimodel for a SPL comprises all the models of the system (i.e. a \texttt{FunctionalModel} and other \texttt{ViewpointModels}) and defines a \texttt{MultimodelRelationship}, \texttt{AppliedTo}, that associates a local feature of the feature model with any \texttt{MultiModelEntity} in another model. Notice that a \texttt{MultiModelEntity} can represent different things in different metamodels. For example, a class can be a \texttt{MultiModelEntity} in a certain viewpoint of the system, but a method can also be a \texttt{MultiModelEntity} in a different viewpoint model. Notice also that, since \texttt{AppliedTo} inherits from \texttt{MultiModelRelationship}, it can only connect sources and targets which are both \texttt{MultiModelEntities}. Thus, we incorporated to the multimodel the class \texttt{FeatureModelMultimodel}, which just inherits from \texttt{FeatureModel}. In \cref{fig:multimodel-metamodel-local-features}, it can also be seen that 
a \texttt{FunctionalModel} is composed of many \texttt{FeatureModel}. One of them is the global feature model, and the other ones are the local feature models.

\begin{figure*}
  \centering
  \includegraphics[width=0.9\linewidth]{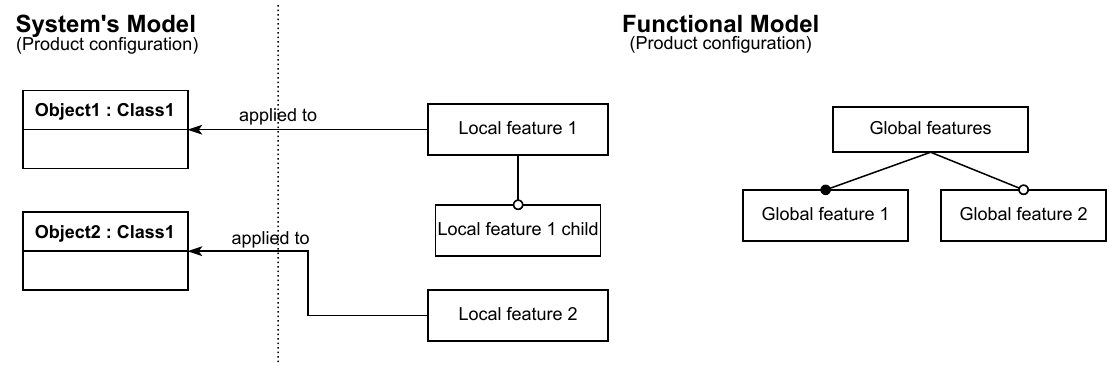}
  \caption{Definition of local features in SPL.}
  \label{fig:definition-local-features}
\end{figure*}

Throughout the article, we will explain with examples the existing relationships between different models with the feature model for the application of local features. To illustrate these relationships, we will use columns to represent different models of the system, and we will use arrows with the annotation \textit{applied to} to indicate the association of features with elements of the other models. \Cref{fig:definition-local-features} shows a simple example. The figure illustrates the product configuration of an unspecified \textit{system model} with two objects of the same class on the left column. On the right column, it depicts the functional model consisting of two local features that apply only to those objects (\textit{Local feature 1} and \textit{Local feature 2}) and a global feature model with features that affect the complete system (\textit{Global features}).
In this \addedtextRR{\st{concrete}} example, you can observe that for \textit{Object1}, only \textit{Local feature 1} is applied, whereas for \textit{Object2}, \textit{Local feature 2} is applied. It is important to note that due to space constraints, the tree of local features has not been depicted alongside the tree of global features. However, these local features would also be present within the global feature tree to enable the definition of global behaviors. Additionally, this setup allows the overwriting of specific behaviors only for the objects of interest. The behavior of these local features will be discussed in more detail below with concrete examples.

\subsection{Example}

In order to see how local features are applied in practice, the following is a case example of an SPL oriented to the generation of e-commerce stores. The applications generated by the SPL are intended for the sale of products that can be very varied: from digital products, such as films, songs, books, etc., to physical products, such as pencils or pens. These products can be found sorted by categories and even subcategories. Payment can be made online and, finally, these applications will have a product catalog where products are grouped by categories.

\begin{figure*}
  \centering
  \includegraphics[width=\linewidth]{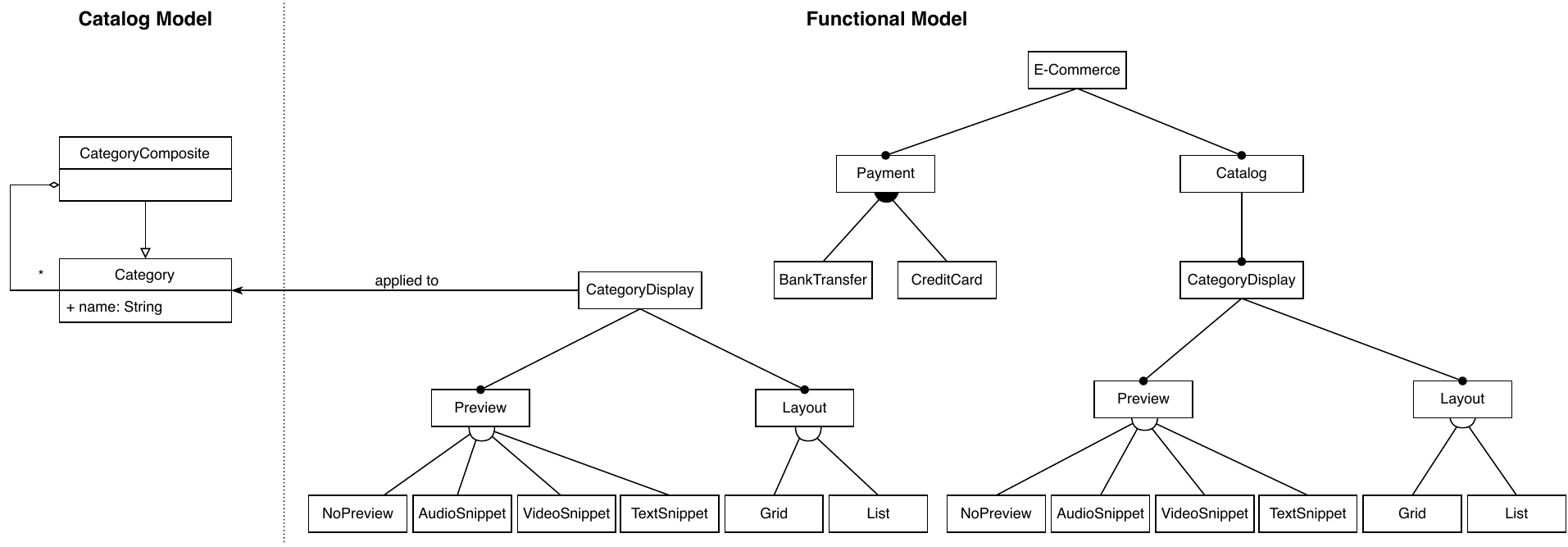}
  \caption{Sample SPL model for the generation of e-commerce applications.}
  \label{fig:proposal-example-spl-model}
\end{figure*}

\Cref{fig:proposal-example-spl-model} shows the multimodel of this SPL, which consists of two models: a catalog model (on the left) and a set of feature models (on the right). The catalog model consists of a class \texttt{CategoryComposite} that inherits from \texttt{Category}, which has only one attribute named \texttt{name}. These two classes represent the categories and subcategories of the products using the composite pattern. Hence, the application engineer can describe the catalog model of an e-commerce store by defining a tree of categories.

The right side of \cref{fig:proposal-example-spl-model} shows the functional model, which contains a global feature model (on the right) and a local feature model (on the left):

\begin{itemize}
    \item \textbf{\texttt{E-Commerce}}: this is the global feature model, i.e. the selection of one of the features in this feature tree affects the whole product. Within this tree are the features related to payment methods (\texttt{Payment}), whose payment can be made by bank transfer (\texttt{BankTransfer}) or by credit card (\texttt{CreditCard}). In turn, there is a second feature, \texttt{Catalog}, which adds the functionality of being able to visualize the products in catalogs grouped by categories. The features within its feature tree are grouped in two subtrees: \texttt{Preview} and \texttt{Layout}. The \texttt{Preview} features sub-tree contains the features related to the preview of the products in the catalog, so depending on the feature selected here the product preview will change. The \texttt{NoPreview} feature does not activate the product preview, \texttt{AudioSnippet}, \texttt{VideoSnippet} and \texttt{TextSnippet} activate the audio, video or text preview, respectively, adding the corresponding audio, video or epub player. The other sub-tree of features, \texttt{Layout}, contains the features associated with how the different products are arranged within a catalog; this can be in the form of a grid (by selecting the \texttt{Grid} feature) or in the form of a list (\texttt{List} feature).
    
    \item \textbf{\texttt{CategoryDisplay}}: this is a local feature model that is a subset of the \texttt{CategoryDisplay} tree of the global feature model seen before, so that each \texttt{Category} class of the data model is related to a \texttt{CategoryDisplay}. Hence, for each category that is created by the application engineer, could have a \texttt{CategoryDisplay} feature tree that defines the specific preview and layout of the products of the category. In the case that this does not exist, the behavior that this \texttt{CategoryDisplay} would be the default one defined in the \texttt{E-Commerce} tree.
\end{itemize}

\begin{figure*}
  \centering
  \includegraphics[width=\linewidth]{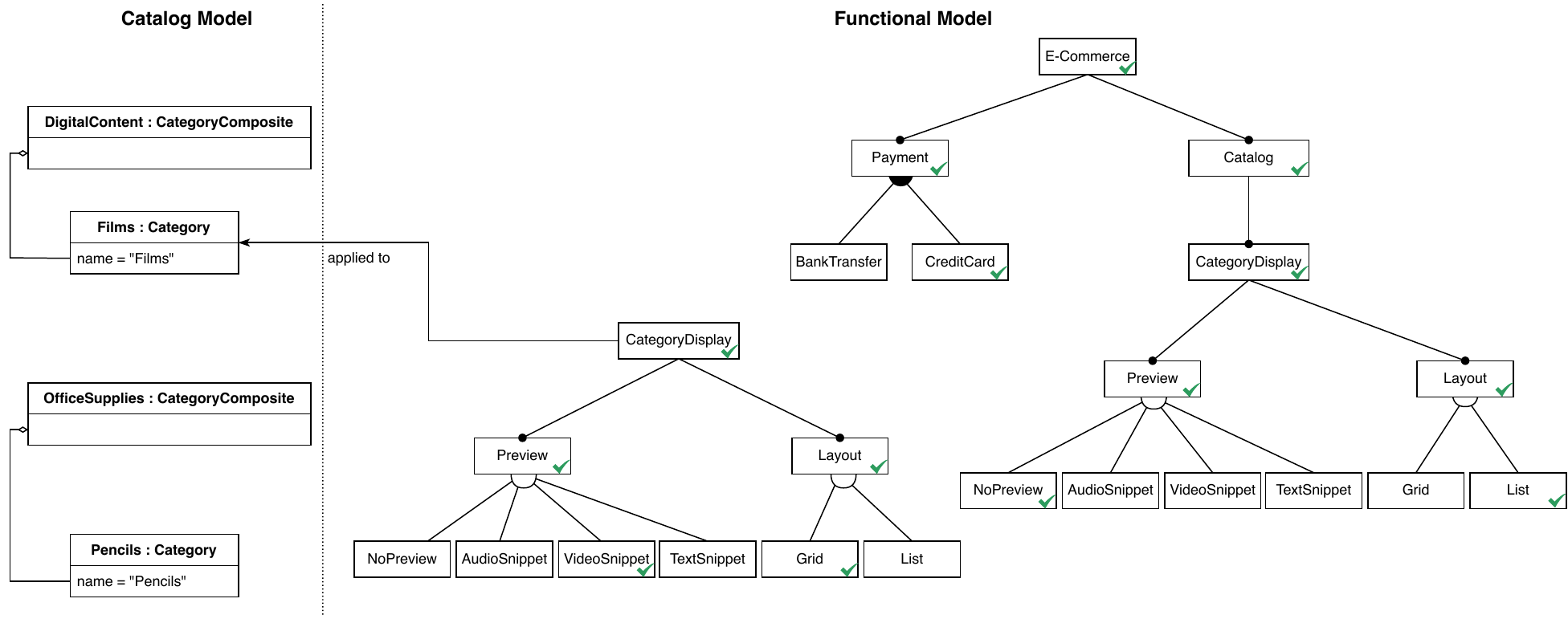}
  \caption{Example of product configuration with the SPL.}
  \label{fig:proposal-example-product-config}
\end{figure*}

In \cref{fig:proposal-example-product-config}, we show an instance of the model in \cref{fig:proposal-example-spl-model} representing the configuration of a specific product. On the left side on the figure, there is the specification of the different categories of the product. On the right side, the chosen functionalities (from the feature models) that the product will have (both global and local ones that apply only to certain categories).
For this example we have defined two main categories, \textit{OfficeSupplies} and \textit{DigitalContent} (instances of the CategoryComposite class) and two subcategories, \textit{Pencils} and \textit{Films} (instances of the Category class). 
The category \textit{Pencils} is not associated with any local feature. Being a physical product, the more suitable way to represent them in a store is through a list and without preview. Since the global feature model already has the default behavior of listing the elements of a category (\texttt{List}) and not having a preview (\texttt{NoPreview}), no local feature has been associated to it to change this behavior.
On the other hand, there is the \textit{Films} category, which, unlike the previous category, can be previewed. Hence, it is associated to the feature \texttt{VideoSnippet} through the applied local feature model with the root feature \texttt{CategoryDisplay}. The feature \texttt{Grid} is also selected for this category, since this type of layout can be more attractive to potential buyers and gives prominence to the preview. 
Finally, there are other global features that affect the entire product; in this case, the feature \texttt{Payment} and its selected subfeature, \texttt{CreditCard} (so you can only pay by card), and the mandatory feature \texttt{Catalog} (the selection of these characteristics is represented in the diagram with green checks to facilitate comprehension). 

By modeling variability in this way, making use of local features, it is possible to have different ways of previewing according to the nature of the product that is for sale. If the model did not have local features, the generated product could not have the preview of the products since pencils are not a video and therefore cannot contain a preview. However, by having local features, it is possible to configure the product as shown in \cref{fig:proposal-example-product-config}. This example shows the practical usefulness of local features and how they improve the level of expressiveness of the feature models. Not only can we choose the features we want our product to have, but we can also specify which specific elements of the product to be generated will have those features.




\section{Case study: developing Geographic Information Systems with local features}
\label{sec:case-study}

In this section, we describe the case study that in fact motivated the need for local features. First, we introduce the context for the case study, and afterward, we describe how our proposal was applied and implemented in our case.

\subsection{Context}\label{subsec:context}

In a previous work, we addressed the development of Geographic Information Systems (GIS) with SPLs \cite{CortinasAlvarez17}. The main characteristic of GIS is that they manage entities with a geospatial component. This characteristic affects all the software layers, from the database, which has to support geospatial data types and operations, to the user interface, which usually presents the information in maps and layers. GIS also support specific functionalities, such as route calculation or data processing based on their spatial representation. These systems are intensively used in public administrations and private companies since they are a typical tool in managing infrastructures (such as supply or transportation networks) or mobility scenarios (such as logistics or mobile workforces), for example. 

Organizations such as ISO\footnote{International Organization for Standardization, \texttt{https://iso.org}} or OGC\footnote{Open Geospatial Consortium, \texttt{https://www.ogc.org}} have made an important standardization effort in GIS. The result is that most GIS have the same software architecture and share tools, libraries, and software components in their development, independently of their application domain. Furthermore, GIS in different application domains share many functionalities. This was the motivation to explore the application of an SPL approach to GIS: our SPL comprises the features that may appear in a GIS and allows the engineer to select which of them must be included in a specific product.

At the Databases Laboratory\footnote{Databases Laboratory website: \url{https://lbd.udc.es/}.}, we have been working on a Software Product Line that generates web-based GIS applications; that is, web applications that allow users to visualize and interact with geographic data, mainly through maps, as well as to offer other functionalities that were identified as common for GIS after an analysis of the domain~\cite{CortinasAlvarez17}. This SPL has been employed in the industry for developing several products, including heritage management, facility management, reduced-mobility accessibility, and public transport management. Some of these products are large-scale, such as WebEIEL\footnote{\url{https://webeiel.dacoruna.gal}}. This is a product family with similar functionalities such as map viewing and user location. However, managing the functional variability of different products is not enough in the domain of GIS. The main difference when developing two GIS appears in the model of the domain. For example, a GIS product for a parcel company requires the system to manage the drivers, warehouses, and roads. In contrast, a GIS oriented to promote tourism in a region requires managing hotels, attractions, or natural landscapes. The data model or domain of a GIS affects its whole functionality, starting with the information that is shown in the map viewers and the way this information is drawn. Therefore, the architecture of our SPL is described by three models: a feature model, a data model, and a visualization model. The variability of functionalities is modeled as a feature model, through which the domain engineer can select which of them are required for each of the GIS products to generate. The data model is defined as an UML class diagram that describes the entities, properties and relationships between entities for the product. Finally, the visualization model allows configuring the different maps that the application will have, as well as the layers and styles with which they will be represented. 


During these projects, it was observed that while time-to-market was significantly reduced, the resulting products always presented a common issue. The functionality defined through feature selection in the feature model is consistently applied in a general manner, without the ability to customize it for specific parts of the system. For example, if a functionality affecting the map viewer of the application was selected, it would be applied in the same way across all maps in the generated product. However, some features may need to be applied only to certain elements of the system. For instance, a GIS generated with the SPL would have different maps, and certain features would be necessary for some maps but not for others. Current feature modeling options do not account for this, requiring additional adaptation efforts for each product that software developers must carry out, resulting in a separation of the product from the product family. Therefore, it is necessary to be able to specify features that affect only a specific part of a system.

\begin{figure}[htbp]
  \centering
  \includegraphics[width=0.8\linewidth]{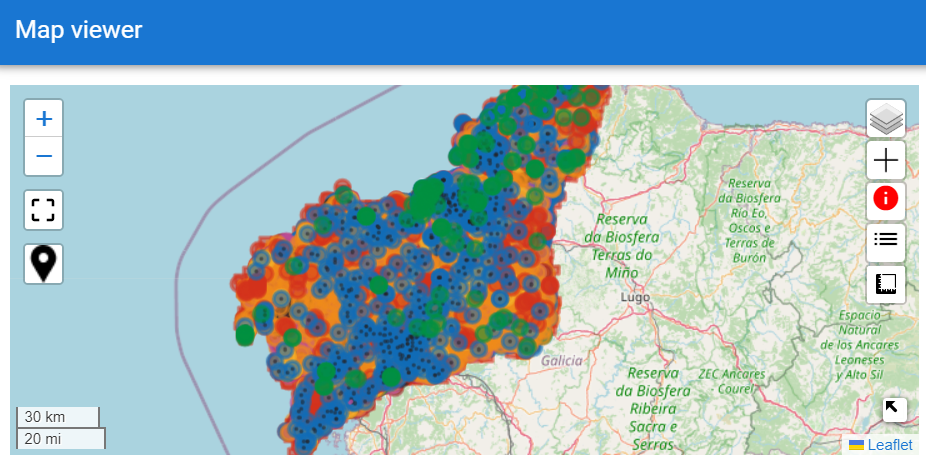}
  \caption{Example of a product generated with a single map viewer}
  \label{fig:example-unique-map-viewer}
\end{figure}

Web\-EIEL\footnote{WebEIEL website: \url{https://webeiel.dacoruna.gal}} is a web application dedicated to publishing geographic information collected by the Provincial Council of A Coruña, Spain, that we will use a a paradigmatic example. Most of the data handled by WebEIEL has a geographic dimension, and it is shown to the user via map viewers with different layers (each layer represents one entity), styled depending on the data represented. There are different types of map viewers. For example, sometimes the application shows a full-fledged map viewer that displays a large collection of layers and provides complex functionality to the user; and sometimes the application requires just a simple map viewer that displays one or few layers and provides almost no functionality to the user beyond its visualization.

WebEIEL's data model is composed of entities that describe infrastructures (e.g., road network) and facilities (e.g., hotels, parks, etc.) related to the municipality where they are located.
The data model is not very complex since most of the relationships are spatial (they do not require foreign keys in the database) and, at the same time, it is extensive because it includes around a hundred entities.
Among the entities, there are different types regarding the functionalities of the application. For example, there are entities that should not be modified, such as the \textit{municipalities}, whereas other entities need to be created and edited from the application, such as \textit{water treatment plants} or \textit{hotels}.



\begin{figure*}[tbp]
  \centering
  \includegraphics[width=\linewidth]{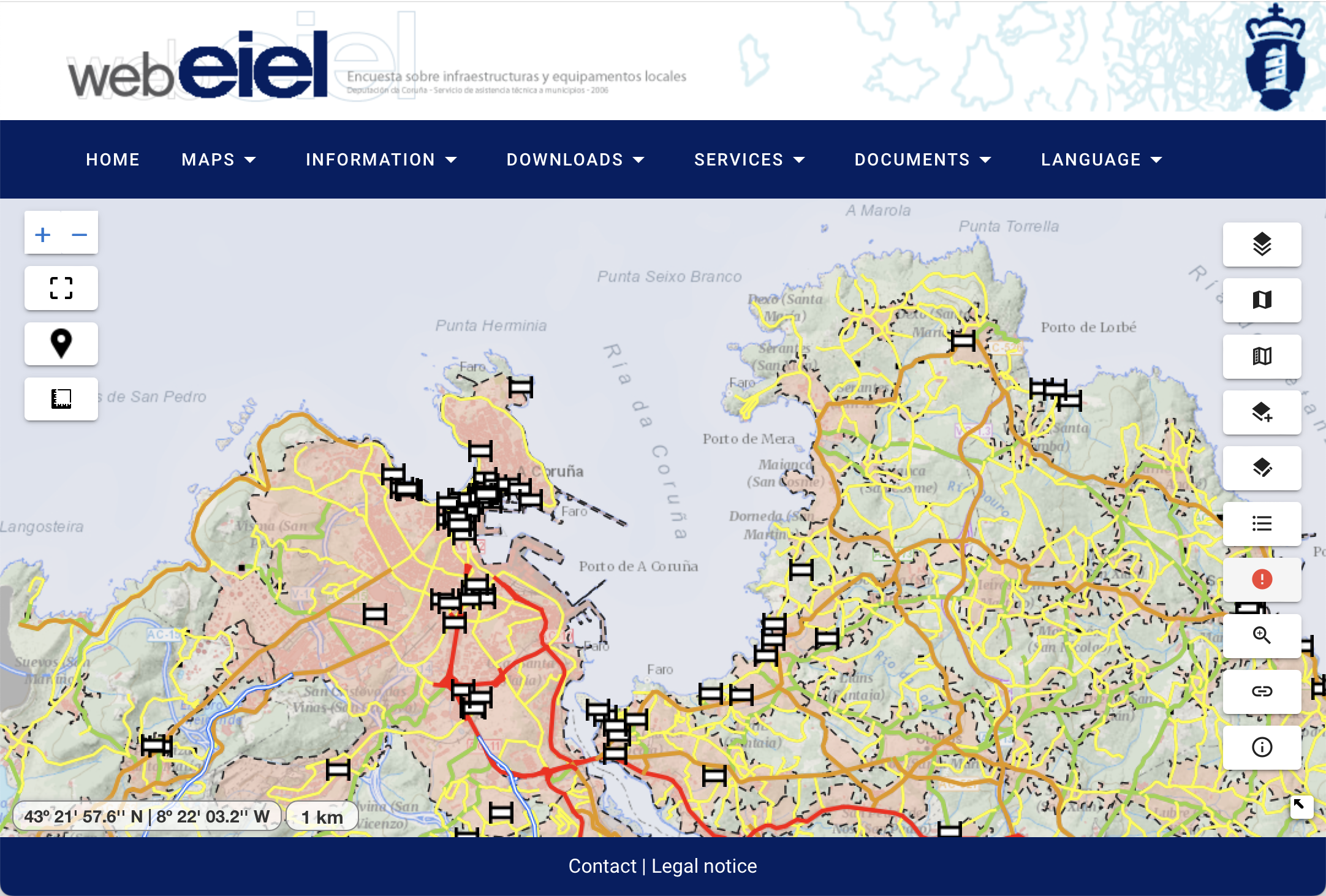}
  \caption{WebEIEL screenshot}
  \label{fig:captura}
\end{figure*}

The first version of WebEIEL was developed in 2008-2010, and was operational until December 2022. It has been technologically outdated for a long time, as resources to its maintenance were scarce.
In 2022, we started a project to upgrade WebEIEL to use current technology, using a previously implemented Software Product Line for the generation of web-based Geographic Information Systems~\cite{CortinasAlvarez17}, evolved in a posterior work which describes the DSL used to generate products of this SPL~\cite{HenrandezAlvarado20}. The version of WebEIEL currently deployed has been generated by the SPL, and modified afterwards by a development team. An example of a map viewer, a concept used all over the section, can be seen in \cref{fig:captura}, and accessed in the actual application\footnote{Example of map viewer showing hotels, among other layers, in WebEIEL website: \url{https://webeiel.dacoruna.gal/map-viewer/movilidad?hash=1677844039765}}.

\begin{figure*}[tbp]
  \centering
  \includegraphics[width=\linewidth]{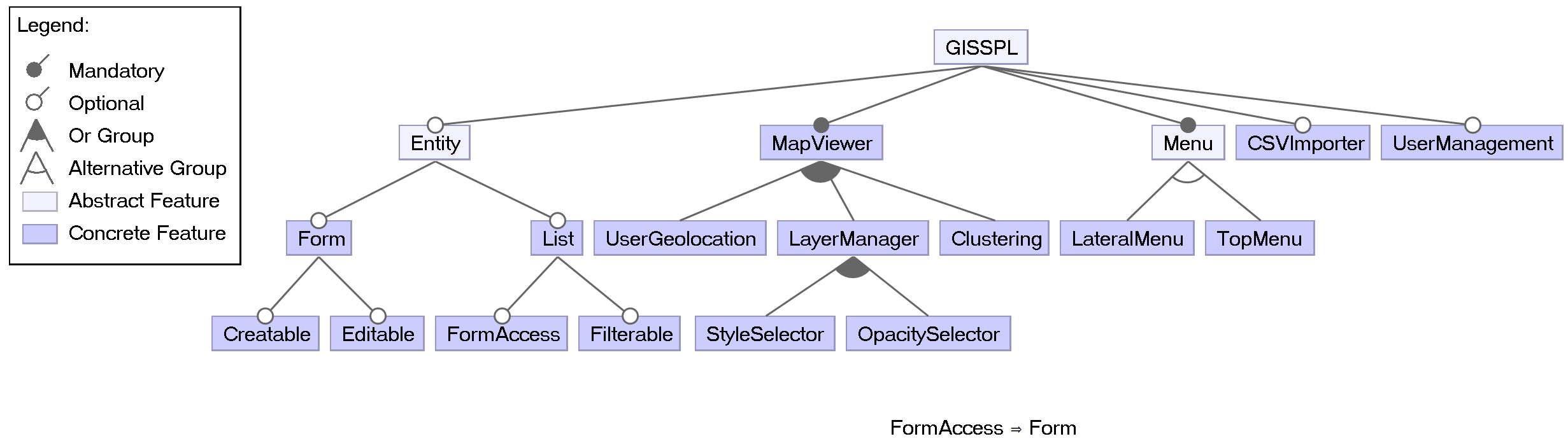}
  \caption{Product Line Feature Model (simplified excerpt, only 18 of the 175 features of our FM are displayed)}
  \label{fig:feature-model}
\end{figure*}

\begin{figure*}[tbp]
  \centering
  \includegraphics[width=0.65\linewidth]{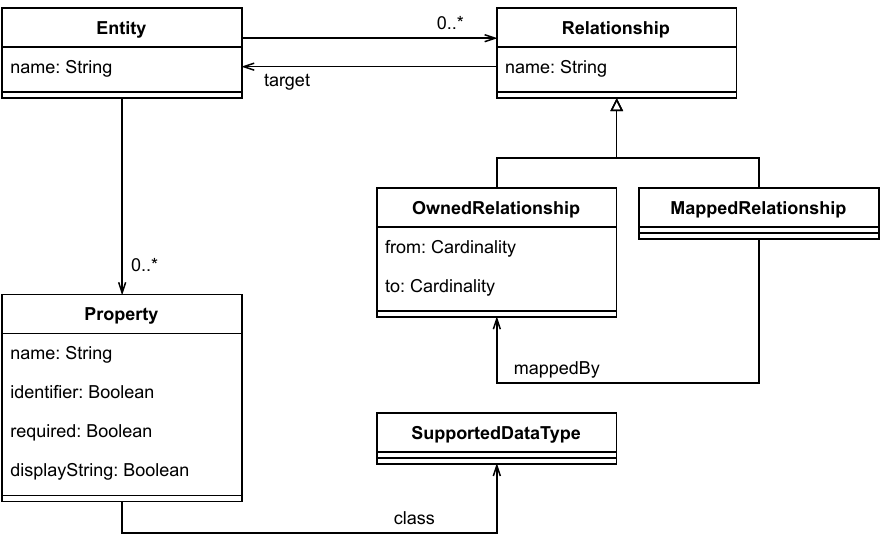}
  \caption{Product Line Data Metamodel}
  \label{fig:data-model}
\end{figure*}

\begin{figure*}[tbp]
  \centering
  \includegraphics[width=0.6\linewidth]{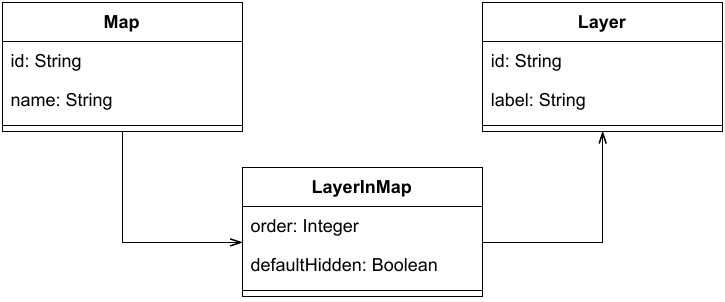}
  \caption{Product Line Visualization Metamodel (simplified excerpt)}
  \label{fig:visualization-model}
\end{figure*}

Briefly explained, our SPL generates web-based GIS that can be specified using three models: a feature model (see \cref{fig:feature-model}), used to determine the functionalities of a product; a data model (see \cref{fig:data-model}), that allows the application engineer to define the entities and relationships of the domain of a specific product; and a visualization model (see \cref{fig:visualization-model}), so the application engineer can configure a product to have different map viewers, each one showing different layers with data. By creating multiple maps, the application engineer can categorize information and tailor the maps' content to specific user groups, making them more appealing and user-oriented. For instance, a map designed for tourists might display only hotels and parks, while another map tailored for a city council worker could show the running water networks. Moreover, this approach significantly enhances the application's performance by avoiding the slowdown caused by loading too many layers simultaneously.

In the feature model (see \cref{fig:feature-model}), the left-most branch defines features related to the entities defined in the data model (\texttt{Entity}). \texttt{Form} generates detailed views of the entities. Depending on the sub-features \texttt{Creatable} and \texttt{Editable}, these detailed views provide the possibility of creating new elements, or editing existing ones. 
The feature \texttt{List} generates listings for each entity, in which all of its elements are displayed in a paginated table. These listings may allow the user to access the detail view of the elements (feature \texttt{FormAccess}, which implies the feature \texttt{Form}), and may be filterable (feature \texttt{Filterable}).

The second branch defines features associated with the maps (\texttt{Map\-Vie\-wer}). The sub-features shown activate functionalities such as geolocating the user (\texttt{User\-Geo\-lo\-ca\-tion}), grouping the geographic elements of a layer by their location (\texttt{Clus\-te\-ring}), and managing the map layers (\texttt{La\-yer\-Ma\-nager}). The latter allows the user to hide and show a specific layer of the map, and also includes two sub-features: one to switch the style of a layer (\texttt{Style\-Selector}), and other to change the opacity of a layer (\texttt{Opacity\-Selector}). 

The last branches define features to decide if the application has a horizontal menu in the top or a left sided vertical menu (\texttt{Menu} and its children); to include a CSV importer that allows loading data into the database from a CSV file (\texttt{CSVImporter}); and to handle user registration and authentication (\texttt{UserManagement}).

The data model (see \cref{fig:data-model}) is one of the three models of our multimodel.
It allows the application engineer to specify the domain of a product, and is somehow a simplified version of the \textit{Object Relational Mapping} model, also very similar to UML. The application engineer defines the entities represented by the \texttt{Entity} class, the properties of the entities in the \texttt{Property} class and the relationships between the entities represented with the \texttt{Relationship} class. This difference with respect to ORM/UML is that our model does not provide support for defining everything that is possible to do with ORM/UML, such as inheritance.

Finally, we have the visualization model (see \cref{fig:visualization-model}) where the user can define maps (\texttt{Map}), layers that present different data (\texttt{Layer}), and the relationship between these two elements, that serves to identify which layers belong to each map (\texttt{LayerInMap}). The represented model is a condensed version of the complete model, aimed at simplifying the solution's comprehension. To achieve this simplification, all classes related to styles have been omitted. These classes are responsible for defining how a layer is visualized in a map, including aspects such as color representation and geometry opacity.


\begin{figure*}[tbp]
  \centering
  \includegraphics[width=0.45\linewidth]{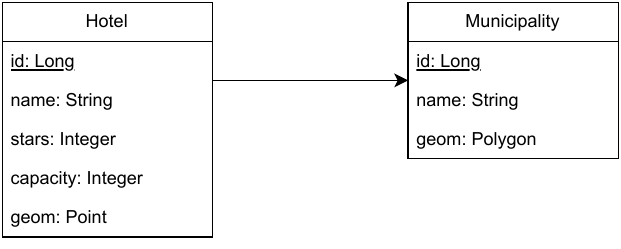}
  \caption{Example of a simple data model for a web-based GIS}
  \label{fig:data-model-instance-example}
\end{figure*}

\begin{figure*}[tbp]
  \centering
  \includegraphics[scale=0.7]{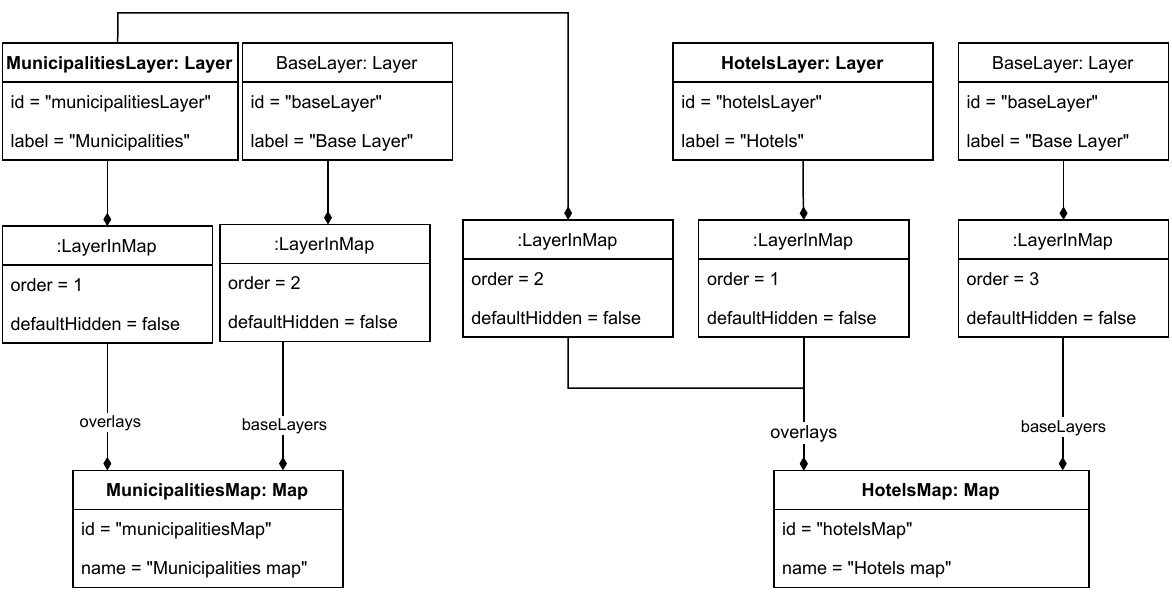}
  \caption{Example of a simple visualization model for a web-based GIS}
  \label{fig:visualization-model-instance-example}
\end{figure*}

Going back to WebEIEL, we have selected, for our example, simplified excerpts 
of the data model, shown in \cref{fig:data-model-instance-example}, and of the visualization model, shown in \cref{fig:visualization-model-instance-example}. The data model contains two entities, \textit{Municipality} and \textit{Hotel}, whereas the visualization model contains two maps: \textit{MunicipalityMap}, that contains a layer to represent municipalities (\textit{MunicipalityLayer}), and \textit{HotelsMap} which contains a layer that represents hotels (\textit{HotelsLayer}), and a layer that represents municipalities (\textit{MunicipalityLayer}). Both maps also contain a base layer (\textit{BaseLayer}) that will work as map background. 

We have already mentioned above that the requirements for the entity \textit{Municipality} are different from the ones for the entity \textit{Hotel}: we do not want the users of our product to create, edit nor remove \textit{municipalities}, but they can do that for \textit{hotels}. Therefore, since at least one of the entities for the product require these functionalities, we need to select the features \texttt{Creatable} and \texttt{Editable} (see \cref{fig:feature-model}). The problem is that these functionalities will be affecting all the entities of the data model, and then a user would be able to modify a \textit{municipality}. To prevent this, a developer needs to modify the generated source code of the product and \textit{remove} functionalities for some of the entities. 

Let us assume that the \textit{MunicipalityMap}, since it only has one overlay, do not require a \textit{layer manager}, whereas the \textit{HotelsMap}, having two overlays, requires a \textit{layer manager}. This is the same case than before: since at least one of the map viewers needs to have a \textit{layer manager}, the analyst has to select the feature \texttt{LayerManager}, and a developer needs to remove the functionality after the product is generated.

Obviously, having to modify the generated products straight after the generation only to remove features is far from ideal, since it rises the time to market, the maintenance costs, and the evolution complexity for the whole product family of the product. In the next section we explain how we adopted \textit{local features} to solve the problems described, showing the changes made to the design of the product line, and also how we approached the implementation.

\subsection{Developed solution}
\label{subsect:solution}

\begin{figure*}[tbp]
  \centering
  \includegraphics[width=\linewidth]{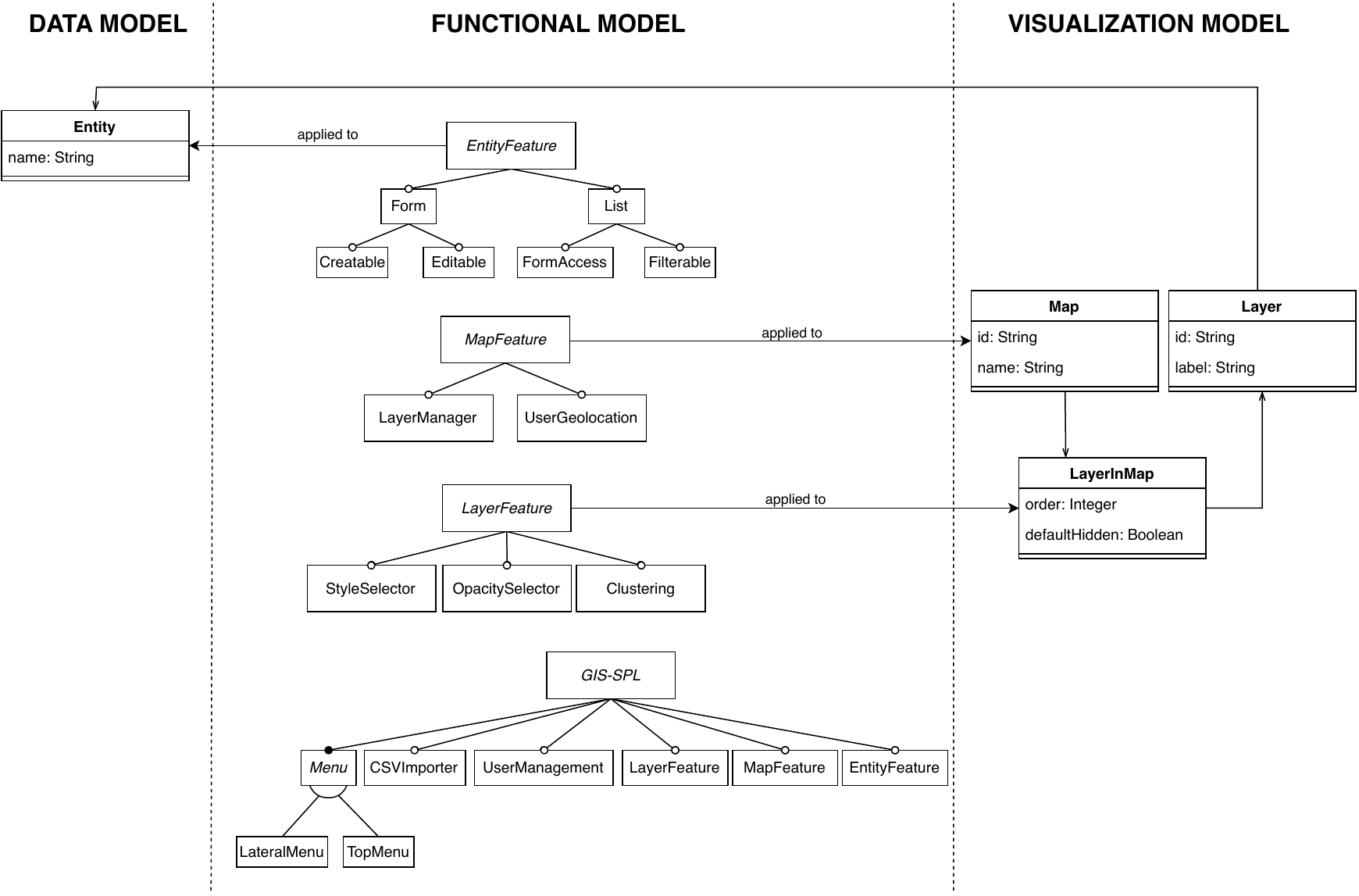}
  \caption{Diagram representing the integration of the features with the concrete elements of the application}
  \label{fig:class-diagram-with-features}
 \end{figure*}

First step is to identify \textit{local features}. We can do that by splitting the original feature model (see \cref{fig:feature-model}) into pieces. The result is shown in \cref{fig:class-diagram-with-features}, where we can see four different feature models. One of them, \texttt{GIS-SPL}, is a global features model and it is not linked to any element, while the other three root features can be \textit{applied to} specific elements of the other models, so they are local feature models (we omitted all the elements which have no local features applied).
Thus, the \texttt{EntityFeature} is applied to the \texttt{Entity} class of the data model, whereas the \texttt{MapFeature} and the \texttt{LayerFeature} are applied to the \texttt{Map} the \texttt{LayerInMap} classes of the visualization model, respectively. 
Hence, when instantiating each of the linked elements, a subset of the local features can be selected, and they will affect only and exclusively the element in question. 
This association can be understood from two points of view. 
From the point of view of the model element, the association represents the functionality available in the product for that specific element. 
For example, if the entity \texttt{Municipality} is associated with the feature \texttt{List}, it means that this entity can be browsed in a list of the product. 
From the point of view of the feature, the association represents the specific configuration of the functionality. 
For example, if the feature \texttt{LayerManager} is associated with an specific instance of a \texttt{Map} entity, it represents that specific map will have a layer manager. 
In case the element does not have a specific configuration associated with it, the default configuration will be the one specified in the global feature model. That is why in \cref{fig:class-diagram-with-features}, within the global feature model, the features \texttt{EntityFeature}, \texttt{MapFeature} and \texttt{LayerFeature} are repeated (for space reasons their subtrees are not shown in the figure, being identical to the local feature models above them).

As this new mechanism provides more expressiveness to the SPL, it is possible to modify the initial feature model of the SPL to take advantage of the additional expressiveness. For example, the \texttt{Clustering} feature in \cref{fig:feature-model} is a child of the \texttt{MapViewer} feature because it was not possible to represent that a map is composed of layers, and that some of them may be clustered if needed. With the new mechanism, we have enough expressiveness to be able to indicate that the \texttt{Clustering} feature is applied only to certain layers, and therefore the feature is a child of the feature \texttt{LayerFeature}. In addition, some features from the original feature model have been grouped depending on the elements they affect. This is why new features appear, such as \texttt{EntityFeature}, \texttt{LayerFeature} and \texttt{MapFeature} that group the functionality of entities, layers and maps respectively.

\begin{figure*}[tbp]
  \centering
  \includegraphics[width=0.8\linewidth]{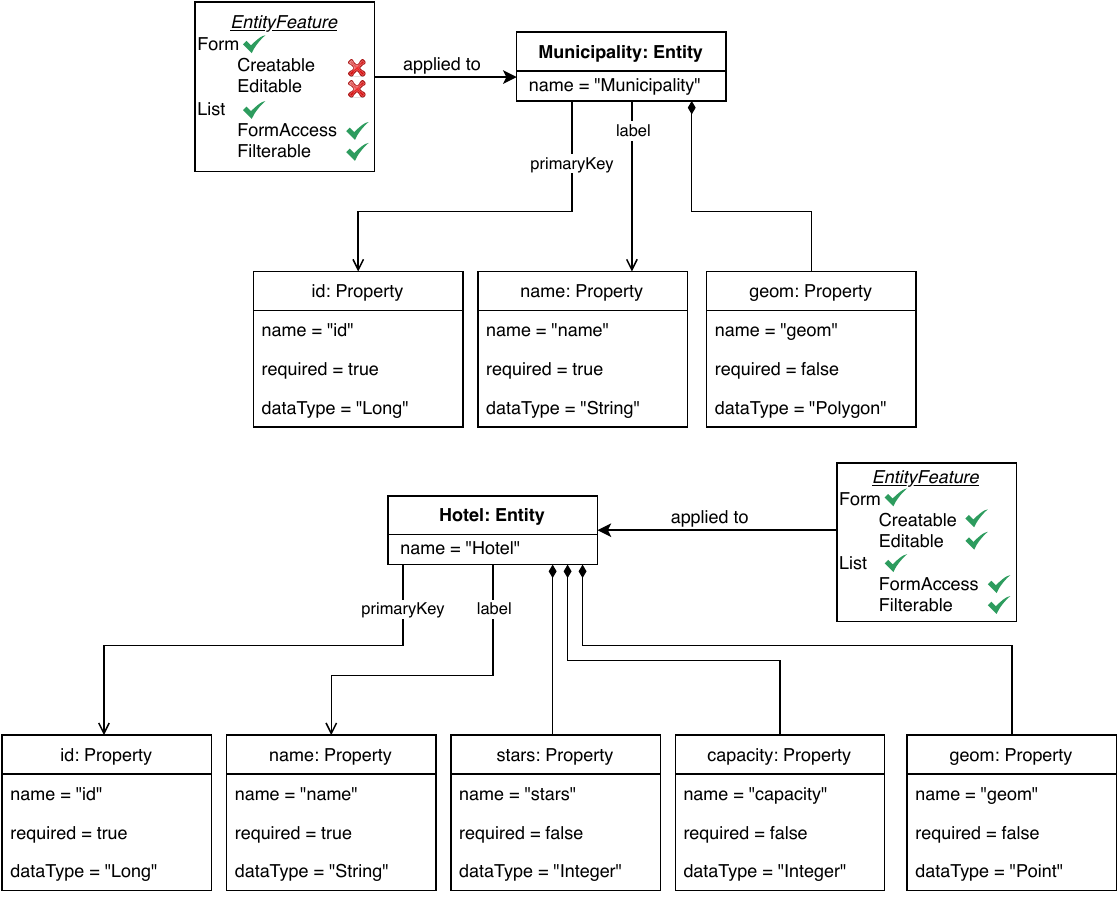}
  \caption{Object diagram of the entities}
  \label{fig:class-diagram-entities}
\end{figure*}

Taking this paradigm change into account, let us follow with the examples described in \cref{subsec:context}. First of all, we have two different entities, \textit{Municipality} and \textit{Hotel}, which require different features. Both entities have a number of properties such as \textit{id} or \textit{name}. Some of these properties are marked as required, so the user who creates objects of these entities must specify at least those marked as required. The decision of which properties are required and which are not is up to the domain engineer and, in this example, we have decided to mark as required only those that we consider essential when creating an object, such as its identifier and name. In addition, both entities are related to each other since a hotel is located within a municipality. \Cref{fig:class-diagram-entities} shows an object diagram that defines the entities of the example and the features associated to each entity from the feature model defined in \cref{fig:class-diagram-with-features}. We can see how the \textit{Municipality} entity is linked to a instance of the \textit{local feature} \texttt{EntityFeature}, and some of its sub-features are selected (\texttt{Form}, \texttt{List}, \texttt{FormAccess}, and \texttt{Filterable}), while \texttt{Creatable} and \texttt{Editable} are not selected. However, we can see that the entity \texttt{Hotel} has all the sub-features of the \textit{local feature} \texttt{EntityFeature}.

\begin{figure}[htbp]
  \centering
  \includegraphics[width=0.6\linewidth]{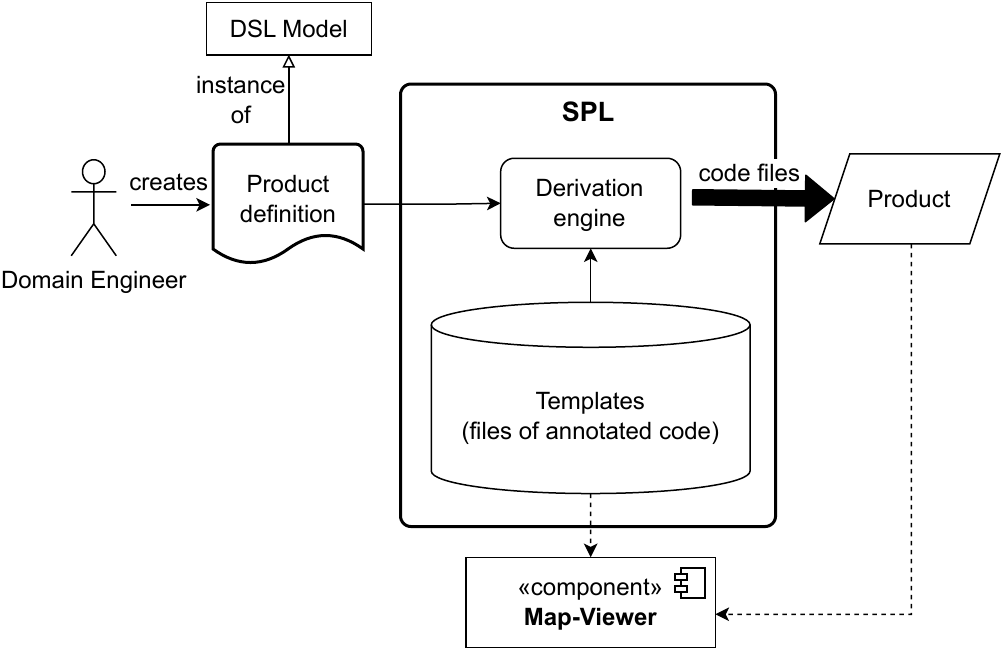}
  \caption{Architecture diagram}
  \label{fig:architecture-diagram}
\end{figure}

In order to define these models, we have used a Domain Specific Language (DSL), which evolves from the one used in \cite{HenrandezAlvarado20}, that allow us to define the data model, the visualization model, and the local features of the elements that may have one. The DSL was specified with a BNF grammar and its parser was implemented with ANTLR\footnote{\texttt{www.antlr.org}}. The resulting architecture is as shown in  \cref{fig:architecture-diagram}. The SPL presented in this application case is based on an annotative derivation engine we developed in a previous project, and which was presented at SPLC'22 \cite{spl-js-engine}. The derivation engine receives a JSON file with the specification of the configurations that have to be applied to the source code templates and generates the final product. This derivation engine supports the concept of multimodel, since the JSON configuration file must contain the product's feature model, but also other models such as the data model, visualization model, or navigation model (although other types of models could be used in different SPLs). This derivation engine is freely available \footnote{\url{https://github.com/AlexCortinas/spl-js-engine}}. The DSL used for the examples described is shown next. 

\begin{lstlisting}[caption={Definition of entities for the WebEIEL product}, label={lst:webeiel-entities-definition}]
CREATE ENTITY Municipality (
    id Long IDENTIFIER,
    name String DISPLAY_STRING REQUIRED,
    geom Polygon,
    hotels Hotel RELATIONSHIP(1..1, 0..*) BIDIRECTIONAL
) WITH FEATURES (Form, List, FormAccess, Filterable);

CREATE ENTITY Hotel (
    id Long IDENTIFIER,
    name String DISPLAY_STRING REQUIRED,
    stars Integer,
    capacity Integer,
    geom Point,
    municipality Municipality RELATIONSHIP MAPPED_BY hotels
) WITH FEATURES (Form, Creatable, Editable, List, FormAccess, Filterable);
\end{lstlisting}

\Cref{lst:webeiel-entities-definition} shows the definition of the entities \textit{Municipality} (L1) and \textit{Hotel} (L8). For both, a series of properties are defined (such as \textit{id}, \textit{name} or \textit{capacity}) and, at the end of the definition, the features that are associated to each entity are indicated by means of the statement \texttt{WITH FEATURES}.

\begin{figure*}[tbp]
  \centering
  \includegraphics[width=\linewidth]{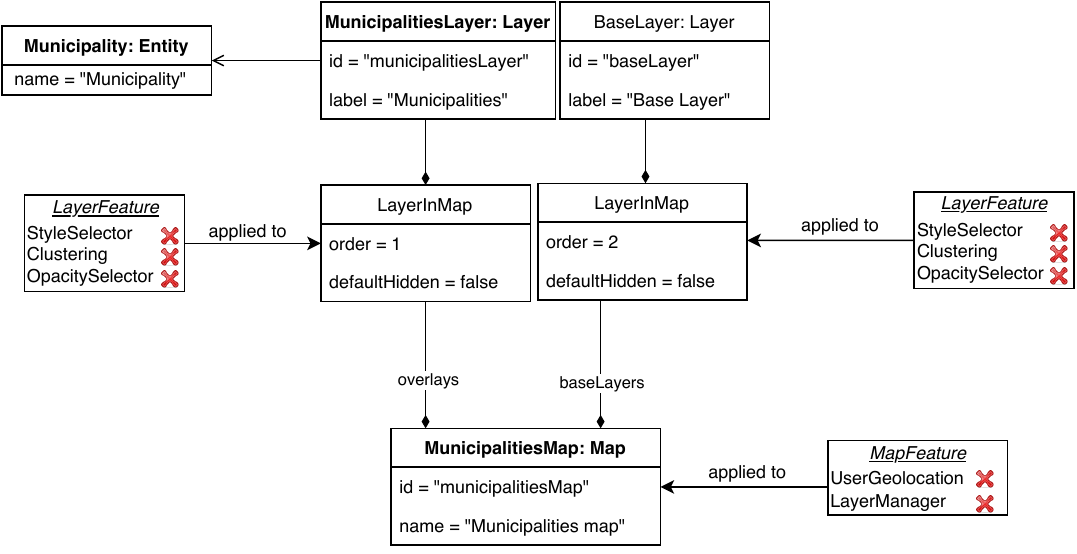}
  \caption{Object diagram of the municipalities map}
  \label{fig:class-diagram-municipalities-map}
\end{figure*}

\begin{figure*}[tbp]
  \centering
  \includegraphics[width=\linewidth]{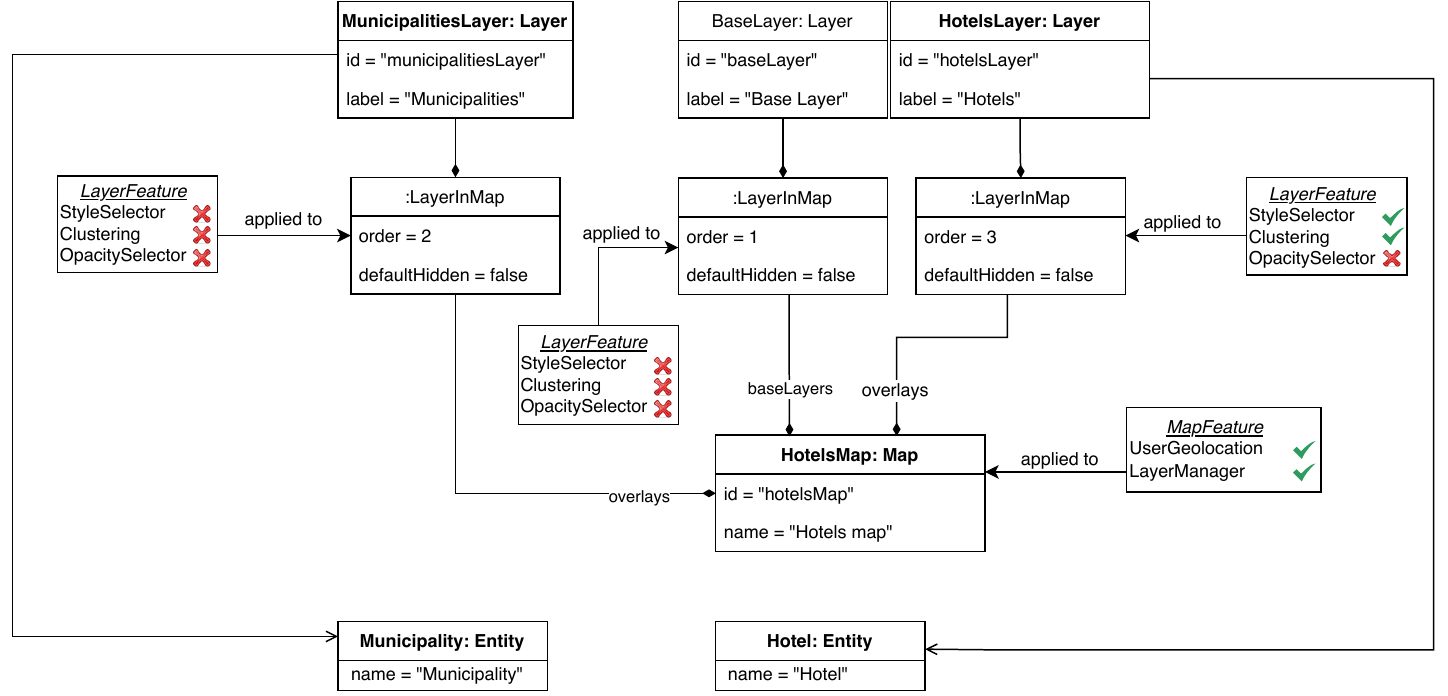}
  \caption{Object diagram of the hotels map}
  \label{fig:class-diagram-hotels-map}
\end{figure*}

The example of the \textit{MunicipalityMap} and the \textit{HotelsMap} can be seen in \cref{fig:class-diagram-municipalities-map} and in \cref{fig:class-diagram-hotels-map}.
The former shows an object diagram that defines a simple map viewer that displays the municipalities. The map contains two layers: the municipalities and a base layer (e.g., the OpenStreetMap\footnote{OpenStreetMaps website: \url{https://www.openstreetmap.org/}} tiles to show the map context). This map viewer requires no features, since it provides few functionality to the user (i.e., the user cannot change the layers and cannot use the geolocation feature). 

\begin{lstlisting}[label={lst:webeiel-municipalitymap-definition}, caption=Definition of the municipalities layer and map]
CREATE GEOJSON LAYER municipalitiesLayer AS Municipalities FOR Municipality
WITH STYLES (
	blueColor DEFAULT
);

CREATE MAP municipalitiesMap AS Municipalities map WITH LAYERS (
    baseLayer IS_BASE_LAYER DEFAULT_BASE_LAYER,
	municipalitiesLayer
), WITH CENTER [ [40.712, -74.227], [40.774, -74.125] ];

\end{lstlisting}

\Cref{lst:webeiel-municipalitymap-definition} shows the definition of the municipalities map in the DSL. It first defines the \textit{municipalitiesLayer} (L1), just setting its name and a style\footnote{The visualization model includes styles for the layers, which have been omitted for clarity in the excerpts and explanations.}. Then, it defines (L6) the \textit{municipalitiesMap} that shows the different municipalities of the province using the layer \textit{municipalitiesLayer}. As we can see, there is not mention to any feature since none is required.

\Cref{fig:class-diagram-hotels-map} shows an object diagram that defines a more complex map viewer, \textit{HotelsMap}, that displays both hotels and municipalities. The map contains three layers: the hotels, the municipalities and a base layer. The map viewer provides much more functionality to the user than \textit{MunicipalitiesMap}. In this case, the user can change the layers, and use the geolocation feature to zoom the map to his/her current position. Similarly, the hotels layer also provides more functionality (i.e., the user can change the style and the geographic objects are clustered to avoid cluttering the map at low scales). 

\begin{lstlisting}[label={lst:webeiel-hotelsmap-definition}, caption=Definition of the hotel layer and map]
CREATE GEOJSON LAYER hotelsLayer AS Hotels FOR Hotel
WITH STYLES (
	starsStyle DEFAULT,
	capacityStyle
);

CREATE MAP hotelsMap AS Hotels map WITH LAYERS (
    baseLayer IS_BASE_LAYER DEFAULT_BASE_LAYER,
    municipalitiesLayer,
    hotelsLayer WITH FEATURES ( StyleSelector, Clustering )
), WITH CENTER [ [40.712, -74.227], [40.774, -74.125] ]
WITH FEATURES ( LayerManager, UserGeolocation );
\end{lstlisting}

\Cref{lst:webeiel-hotelsmap-definition} shows the definition of the hotels map in the DSL. Line 1 it defines the \textit{hotelsLayer} with multiple styles that represent the hotels depending of their number of stars or its capacity (\textit{starsStyle} and \textit{capacityStyle} respectively). Then, the \textit{HotelsMap} is defined containing a base layer, the municipalities, and the hotels (L7). The \textit{hotelsLayer} is linked to the features \texttt{StyleSelector} and \texttt{Clustering} (to be able to switch between its different styles, and to cluster the objects to avoid cluttering the map). Besides that, the \textit{HotelsMap} is linked to the features \texttt{LayerManager} and \texttt{UserGeolocation}. 

\begin{lstlisting}[label={lst:webeiel-product-definition}, caption=Definition of the WebEIEL product with global features]
CREATE GIS WebEIEL WITH FEATURES (TopMenu, UserManagement);
\end{lstlisting}

Finally, we have to define the features that are global to the product. \Cref{lst:webeiel-product-definition} shows an example that selects the \texttt{TopMenu} feture and the \texttt{UserManagement} feature from the feature model shown in \cref{fig:feature-model}.

This case study shows that our proposal allows a much deeper level of customization when defining maps and layers because the features are assigned to each of them individually. 
It is important to remark that this level of customization is possible because some of the features of the feature model have been associated with parts of the product by means of the DSL. Furthermore, in our implementation of the SPL, a product does not include every time all the features and only activates them for the selected entities/elements. The features will only be included if there is an entity or element that has it selected. For example, if no entity of the model has selected the forms feature (i.e., \texttt{Form}), the source code associated with this feature will not be included in the product.

The complete object diagram of the example can be seen in \cref{fig:webeiel-object-diagram} (that can be found in \cref{WebEIEL-object-diagram}). The object diagram represents the result of instantiating entities, maps and layers with the DSL to generate the WebEIEL product. The diagram shows how each element has a series of associated features independent of the others, as it has been explained throughout this section.

\subsection{Discussion}

Next, we present a summary of the application of local features to entities, maps, and layers. The data model of WebEIEL consists of 107 entities:

\begin{itemize}
    \item 17 entities are used to represent indicators computed from the data and they are just pairs of (administrative region, indicator value). Therefore, these entities do not have an associated form (i.e., these entities do not require the feature \texttt{Form}) and they are only listed applying the feature \texttt{Filterable}.

    \item 11 entitites represent the context of the maps (e.g., provinces, municipalities, hidrography, etc.). These entities have an associated form (feature \texttt{Form}), and hence they are listed applying the local features \texttt{FormAccess} and \texttt{Filterable}. However, since their values must not be modified, visualizing them as in a non-editable form was the best decision and the local features \texttt{Creatable} and \texttt{Editable} are not applied.

    \item The other 79 entities have associated forms with the features \texttt{Creatable} and \texttt{Editable}, and are listed with the \texttt{FormAccess} and \texttt{Filterable} features.
\end{itemize}

WebEIEL also includes 54 maps. 44 of them do not need the features \texttt{LayerManager} nor \texttt{UserGeolocation} because they display province-wide indicators with a single layer where geolocating the user is not useful.The remaining 8 maps need the feature \texttt{LayerManager} and \texttt{UserGeolocation} because they use multiple layers and show detailed data. The product has 150 layers. In this case, the feature \texttt{Clustering} was not applied to any of those layers, the feature \texttt{OpacitySelector} was applied to all of them, and the \texttt{StyleSelector} was applied to some of the layers depending on the particular needs of each case.

These numbers show the high degree of customization that was necessary to apply to the entities, maps, and layers of the system. The case study shows how the concept of local features was applied in a real, non-trivial SPL. The description of the case study shows that the concept of local feature appears in many parts of this SPL, and the resulting solution shows the feasibility and adequacy of the concept of local features for modelling and specifying variability in systems where some features must be bound to specific elements of the system specified in other system's models. As we have explained in the study of alternatives presented in the introduction, addressing this requirement with existing variability modelling proposals would result in solutions that we consider unsatisfactory, either because they lead to artificially complex feature models or because they lead to bad design decisions.

\section{Conclusions and Future Work}
\label{sec:conclusions}

SPLs have become a relevant paradigm in the development of families of software products, with many success cases in industrial settings. SPLs reduce development costs and increase product quality by allowing us to semi-automatically develop software products from a set of core assets that are adapted and customized based on the selection of features that must be present in the product we want to generate. In this article, we presented the concept of local feature, which brings the possibility of associating features to the system's elements to which they should be applied in the application engineering phase, that is, when the product to be generated is defined and configured. This allows us to apply features to custom system elements, and not just to pre-established variation points decided at the domain engineering phase. Local features allow us to further customize the products generated with the software platform, which consequently allows us to better meet software requirements, as we have seen in many examples throughout the article.

The article develops the concept of local features to propose a practical implementation. In our proposal, the feature model comprises two types of features: global and local. Global features are either selected or not selected when configuring a product and, if selected, they are applied at a predefined variation point (in the domain engineering phase). In the case of local features, if they are selected for a new product, they must be associated to the system's elements to which they must be applied. Therefore, the variation points affected by a local feature are decided at the application engineering phase, that is, when we define  and build a new product. Since we need to associate local features to other system's elements, we model local features using the concept of multimodel, which allows us to establish associations between features and elements of other system's models, such as the data model. We have provided different examples that illustrate the usefulness of the concept of local features in real scenarios.

In order to demonstrate the usefulness and benefits of this approach, we applied our proposal to an SPL for GIS applications. One of the problems to be solved regarding the implementation of local features is the mechanism to define their associations with other system's elements. We proposed a domain specific-language to define these associations. In the application case, we showed how local features could be used to customize the application of the features regarding data visualization in maps based on the application requirements. The visualization model of a GIS defines which maps the users will see and how the information regarding entities with a geo-spatial component is organized into layers that will be part of those maps. The definition of the visualization model is not standard, that is, each GIS has its own visualization model depending on the requirements (for example, the visualization models would be different for a GIS for road management than that of a GIS for electricity supply management). This allowed us to decide in the application engineering phase which features would be applied to each elements of the visualization model. This application case allowed us to show that our proposal allows us to achieve a higher level of customization of the products generated with the SPL.

The application case we have presented in Section 4 shows how local features were implemented with a DSL. This DSL was implemented as part of a real project and it is, therefore, specific of the domain of GIS (since we considered that in this way it would be easier to use). However, as future work, we are considering the design of a more generic domain specific language that allows us to specify the associations between local features and elements from other system's models.

\section*{Acknowledgements}

This work has been partially funded by the following grants: 0064\_GRESINT\_1\_E partially funded by EU through the Interreg Spain-Portugal / POCTEP; 
PID2022-141027NB-C21 (EarthDL): partially funded by MCIN/AEI/10.13039/501100011033 and EU/ERDF A way of making Europe; 
PID2021-122554OB-C33 (OASSIS): partially funded by MCIN/AEI/10.13039/501100011033 and EU/ERDF A way of making Europe; 
GRC: ED431C 2021/53, partially funded by GAIN/Xunta de Galicia; 
CITIC is funded by the Xunta de Galicia through the collaboration agreement between the Department of Culture, Education, Vocational Training and Universities and the Galician universities for the reinforcement of the research centers of the Galician University System (CIGUS).

\bibliographystyle{elsarticle-num}
\bibliography{library}

\begin{thebibliography}{10}
\expandafter\ifx\csname url\endcsname\relax
  \def\url#1{\texttt{#1}}\fi
\expandafter\ifx\csname urlprefix\endcsname\relax\def\urlprefix{URL }\fi
\expandafter\ifx\csname href\endcsname\relax
  \def\href#1#2{#2} \def\path#1{#1}\fi

\bibitem{weiss99}
D.~M. Weiss, C.~T.~R. Lai, Software Product-Line Engineering - A Family-Based
  Software Development Process, Addison-Wesley, 1999.

\bibitem{Pohl2005}
K.~Pohl, G.~B{\"{o}}ckle, F.~V.~D. Linden, {Software Product Line Engineering:
  Foundations, Principles and Techniques}, Springer, 2005.
\newblock \href {https://doi.org/10.1007/3-540-28901-1}
  {\path{doi:10.1007/3-540-28901-1}}.

\bibitem{Weiss2006}
D.~M. Weiss, P.~Clements, C.~W. Krueger, {Software Product Line Hall of Fame},
  in: Proceedings of the 10th International Software Product Line Conference
  (SPLC'06), 2006, pp. 237--237.
\newblock \href {https://doi.org/10.1109/SPLINE.2006.1691614}
  {\path{doi:10.1109/SPLINE.2006.1691614}}.

\bibitem{boeing-hall-of-fame}
D.~Sharp, Reducing avionics software cost through component based product line
  development, in: Proceedings of the 17th DASC. AIAA/IEEE/SAE. Digital
  Avionics Systems Conference, Vol.~2, 1998, pp. G32/1--G32/8.
\newblock \href {https://doi.org/10.1109/DASC.1998.739846}
  {\path{doi:10.1109/DASC.1998.739846}}.

\bibitem{Iglesias201964}
A.~Iglesias, M.~Iglesias-Urkia, B.~López-Davalillo, S.~Charramendieta,
  A.~Urbieta, Trilateral: Software product line based multidomain iot artifact
  generation for industrial cps, in: Proceedings of the 7th International
  Conference on Model-Driven Engineering and Software Development
  (MODELSWARD'19), 2019, pp. 64--73.
\newblock \href {https://doi.org/10.5220/0007343500640073}
  {\path{doi:10.5220/0007343500640073}}.

\bibitem{Trujillo2007a}
S.~Trujillo, D.~Batory, O.~Diaz, Feature oriented model driven development: A
  case study for portlets, in: Proceedings of the 29th International Conference
  on Software Engineering (ICSE'07), 2007, pp. 44--53.
\newblock \href {https://doi.org/10.1109/ICSE.2007.36}
  {\path{doi:10.1109/ICSE.2007.36}}.

\bibitem{Rincon201571}
L.~Rincon, G.~Rodriguez, J.~C. Martinez, G.~I. Alvarez, M.~C. Pabon, Creating
  virtual stores using software product lines: An application case, in:
  Proceedings of the 10th Computing Colombian Conference (CCC'15), 2015, pp.
  71--78.
\newblock \href {https://doi.org/10.1109/ColumbianCC.2015.7333414}
  {\path{doi:10.1109/ColumbianCC.2015.7333414}}.

\bibitem{apel2009overview}
S.~Apel, C.~K{\"a}stner, An overview of feature-oriented software development.,
  Journal of Object Technology 8~(5) (2009) 49--84.

\bibitem{apel2016feature}
S.~Apel, D.~Batory, C.~K{\"a}stner, G.~Saake, Feature-oriented software product
  lines, Springer, 2016.

\bibitem{Benavides2010}
D.~Benavides, S.~Segura, A.~Ruiz-Cort{\'{e}}s, Automated analysis of feature
  models 20 years later: A literature review, Information Systems 35~(6) (2010)
  615--636.
\newblock \href {https://doi.org/10.1016/j.is.2010.01.001}
  {\path{doi:10.1016/j.is.2010.01.001}}.

\bibitem{Czarnecki2002}
K.~Czarnecki, T.~Bednasch, P.~Unger, U.~Eisenecker, Generative programming for
  embedded software: An industrial experience report, in: Proceedings of the
  International Conference on Generative Programming and Component Engineering
  (GPCE'02), Vol. LNCS 2487, Springer-Verlag, 2002, pp. 156--172.
\newblock \href {https://doi.org/10.1007/3-540-45821-2_10}
  {\path{doi:10.1007/3-540-45821-2_10}}.

\bibitem{Czarnecki2005b}
K.~Czarnecki, S.~Helsen, U.~Eisenecker, {Staged configuration through
  specialization and multilevel configuration of feature models}, Software
  Process Improvement and Practice 10~(2) (2005) 143--169.
\newblock \href {https://doi.org/10.1002/spip.225}
  {\path{doi:10.1002/spip.225}}.

\bibitem{Benavides2005}
D.~Benavides, P.~Trinidad, A.~Ruiz-Cort{\'{e}}s, {Automated Reasoning on
  Feature Models}, in: Proceedings of the 17th International Conference on
  Advanced Information Systems Engineering (CAiSE'05), 2005, pp. 491--503.
\newblock \href {https://doi.org/10.1007/11431855_34}
  {\path{doi:10.1007/11431855_34}}.

\bibitem{Batory2006}
D.~Batory, D.~Benavides, A.~Ruiz-Cortes, {Automated analysis of feature models:
  Challenges ahead}, Communications of the ACM 49~(12) (2006) 2--3.
\newblock \href {https://doi.org/10.1145/1183236.1183264}
  {\path{doi:10.1145/1183236.1183264}}.

\bibitem{6030048}
M.~Voelter, E.~Visser, Product line engineering using domain-specific
  languages, in: Proceedings of the 15th International Software Product Line
  Conference (SPLC'11), 2011, pp. 70--79.
\newblock \href {https://doi.org/10.1109/SPLC.2011.25}
  {\path{doi:10.1109/SPLC.2011.25}}.

\bibitem{Riebisch2002}
M.~Riebisch, K.~B{\"{o}}llert, D.~Streitferdt, I.~Philippow, Extending feature
  diagrams with uml multiplicities, in: Proceedings of the 6th World Conference
  on Integrated Design \& Process Technology (IDPT2002), 2002, pp. 1--7.

\bibitem{Czarnecki2004}
K.~Czarnecki, S.~Helsen, U.~Eisenecker, Staged configuration using feature
  models, in: Proceedings of the International Conference on Generative
  Programming and Component Engineering (GPCE'04), Vol. LNCS 3154, Springer,
  2004, pp. 266--283.
\newblock \href {https://doi.org/10.1007/978-3-540-28630-1_17}
  {\path{doi:10.1007/978-3-540-28630-1_17}}.

\bibitem{Czarnecki2005}
K.~Czarnecki, S.~Helsen, U.~Eisenecker, Formalizing cardinality-based feature
  models and their specialization, Software Process: Improvement and Practice
  10~(1) (2005) 7--29.
\newblock \href {https://doi.org/10.1002/spip.213}
  {\path{doi:10.1002/spip.213}}.

\bibitem{siegmund2020configuration}
N.~Siegmund, N.~Ruckel, J.~Siegmund,
  \href{https://doi.org/10.1145/3368089.3409675}{Dimensions of software
  configuration: On the configuration context in modern software development},
  in: Procs. of the 28th ACM Joint Meeting on European Software Engineering
  Conference and Symposium on the Foundations of Software Engineering (ESEC/FSE
  2020), ACM Press, 2020, p. 338–349.
\newline\urlprefix\url{https://doi.org/10.1145/3368089.3409675}

\bibitem{multimodels-quality}
J.~Gonzalez-Huerta, E.~Insfran, S.~Abrahão, Defining and validating a
  multimodel approach for product architecture derivation and improvement, in:
  Proceedings of the International Conference on Model-Driven Engineering
  Languages and Systems (MODELS'13), Vol. LNCS 8107, Springer, 2013, pp.
  388--404.
\newblock \href {https://doi.org/10.1007/978-3-642-41533-3}
  {\path{doi:10.1007/978-3-642-41533-3}}.

\bibitem{decastro22}
D.~de~Castro, A.~Corti\~{n}as, M.~R. Luaces, O.~Pedreira, A.~S. Places,
  Improving the customization of software product lines through the definition
  of local features, in: Proceedings of the 26th ACM International Systems and
  Software Product Line Conference (SPLC'22) - Volume A, ACM, 2022, p.
  199–209.
\newblock \href {https://doi.org/10.1145/3546932.3547006}
  {\path{doi:10.1145/3546932.3547006}}.

\bibitem{Kang1990}
K.~C. Kang, S.~G. Cohen, J.~A. Hess, W.~E. Novak, A.~S. Peterson,
  Feature-oriented domain analysis (foda) feasibility study, Tech. Rep.
  CMU/SEI-90-TR-021, Software Engineering Institute (1990).

\bibitem{Sousa2016}
G.~Sousa, W.~Rudametkin, L.~Duchien, Extending feature models with relative
  cardinalities, in: Proceedings of the 20th International Systems and Software
  Product Line Conference (SPLC'16), 2016, pp. 79--88.
\newblock \href {https://doi.org/10.1145/2934466.2934475}
  {\path{doi:10.1145/2934466.2934475}}.

\bibitem{Galindo2020}
J.~A. Galindo, D.~Benavides, A python framework for the automated analysis of
  feature models: A first step to integrate community efforts, in: Proceedings
  of the 24th ACM International Systems and Software Product Line Conference
  (SPLC'20), Vol.~B, 2020, p. 52–55.
\newblock \href {https://doi.org/10.1145/3382026.3425773}
  {\path{doi:10.1145/3382026.3425773}}.

\bibitem{Apel2009a}
S.~Apel, C.~K{\"{a}}stner, {An overview of feature-oriented software
  development}, Journal of Object Technology 8~(5) (2009) 49--84.
\newblock \href {https://doi.org/10.5381/jot.2009.8.5.c5}
  {\path{doi:10.5381/jot.2009.8.5.c5}}.

\bibitem{Alferez2019}
M.~Alf{\'e}rez, M.~Acher, J.~A. Galindo, B.~Baudry, D.~Benavides, Modeling
  variability in the video domain: language and experience report, Software
  Quality Journal 27~(1) (2019) 307--347.
\newblock \href {https://doi.org/10.1007/s11219-017-9400-8}
  {\path{doi:10.1007/s11219-017-9400-8}}.

\bibitem{karatas2013}
A.~S. Karata{\c{s}}, H.~O{\u{g}}uzt{\"u}z{\"u}n, A.~Do{\u{g}}ru, {From extended
  feature models to constraint logic programming}, Science of Computer
  Programming 78~(12) (2013) 2295--2312.
\newblock \href {https://doi.org/10.1016/j.scico.2012.06.004}
  {\path{doi:10.1016/j.scico.2012.06.004}}.

\bibitem{multimodel-metamodel}
J.~González-Huerta, E.~Insfran, S.~Abrahão, A multimodel for integrating
  quality assessment in model-driven engineering, in: Proceedings of the 8th
  International Conference on the Quality of Information and Communications
  Technology (QUATIC'12), 2012, pp. 251--254.
\newblock \href {https://doi.org/10.1109/QUATIC.2012.14}
  {\path{doi:10.1109/QUATIC.2012.14}}.

\bibitem{multimodels-def}
E.~Barkmeyer, A.~Barnard, P.~Denno, D.~Flater, D.~Libes, M.~Steves, E.~Wallace,
  Concepts for automating systems integration, Tech. Rep. NISTIR 6928, NIST -
  National Institute of Standards and Technology (2003).
\newblock \href {https://doi.org/https://doi.org/10.6028/NIST.IR.6928}
  {\path{doi:https://doi.org/10.6028/NIST.IR.6928}}.

\bibitem{metzger2007disambiguating}
A.~Metzger, K.~Pohl, P.~Heymans, P.-Y. Schobbens, G.~Saval, Disambiguating the
  documentation of variability in software product lines: A separation of
  concerns, formalization and automated analysis, in: 15th IEEE International
  Requirements Engineering Conference (RE 2007), IEEE, 2007, pp. 243--253.

\bibitem{berger2019usage}
T.~Berger, P.~Collet, Usage scenarios for a common feature modeling language,
  in: Proceedings of the 23rd International Systems and Software Product Line
  Conference-Volume B, 2019, pp. 174--181.

\bibitem{CortinasAlvarez17}
A.~Cortiñas, M.~R. Luaces, O.~Pedreira, A.~S. Places, J.~Perez, Web-based
  geographic information systems sple: Domain analysis and experience report,
  in: Proceedings of the 21st International Systems \& Software Product Line
  Conference (SPLC'17), Vol.~1, 2017, pp. 190--194.
\newblock \href {https://doi.org/10.1145/3106195.3106222}
  {\path{doi:10.1145/3106195.3106222}}.

\bibitem{HenrandezAlvarado20}
S.~H. Alvarado, A.~Cortiñas, M.~R. Luaces, O.~Pedreira, A.~S. Places,
  Developing web-based geographic information systems with a dsl: proposal and
  case study, Journal of Web Engineering (2020) 167--194\href
  {https://doi.org/10.13052/jwe1540-9589.1923}
  {\path{doi:10.13052/jwe1540-9589.1923}}.

\bibitem{spl-js-engine}
A.~Corti\~{n}as, M.~R. Luaces, O.~Pedreira,
  \href{https://doi.org/10.1145/3503229.3547035}{Spl-js-engine: A javascript
  tool to implement software product lines}, in: Proceedings of the 26th ACM
  International Systems and Software Product Line Conference - Volume B
  (SPLC'22), ACM Press, 2022, p. 66–69.
\newblock \href {https://doi.org/10.1145/3503229.3547035}
  {\path{doi:10.1145/3503229.3547035}}.
\newline\urlprefix\url{https://doi.org/10.1145/3503229.3547035}

\end{thebibliography}

\clearpage\onecolumn\appendix

\section{WebEIEL object diagram}\label{WebEIEL-object-diagram}

\begin{figure*}[!h]
  \centering
  \includegraphics[width=0.91\textheight,angle=90]{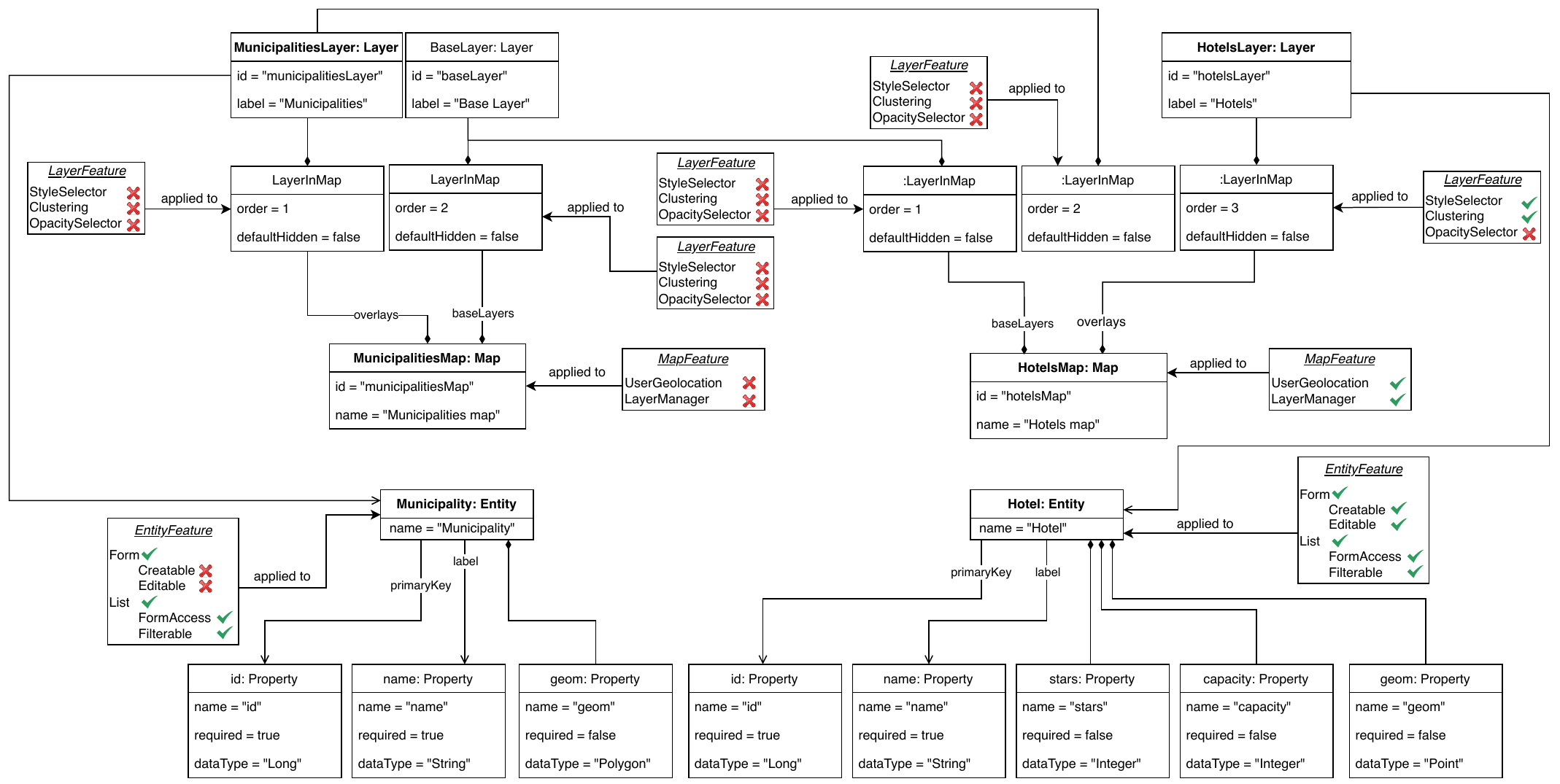}
  \caption{Excerpt of WebEIEL object diagram}
  \label{fig:webeiel-object-diagram}
\end{figure*}

\end{document}